\documentclass{aa} 

\usepackage{graphicx}
\usepackage{txfonts}
\begin{document} 

   \title{Brown dwarfs as ideal candidates for detecting UV aurora outside the Solar
   System: Hubble Space Telescope observations of 2MASS J1237+6526}
   \titlerunning{UV aurora brown dwarf 2MASS J1237+6526}

   \author{Joachim Saur
          \inst{1}
          \and
          Clarissa Willmes \inst{1}
          \and
          Christian Fischer \inst{1}
          \and
          Alexandre Wennmacher \inst{1}
          \and  
          Lorenz Roth\inst{2}
          \and  
          Allison Youngblood\inst{3}
          \and
          Darrell F. Strobel\inst{4}
          \and
          Ansgar Reiners\inst{5}
          }

   \institute{Institut f\"ur Geophysik und Meteorologie, Universit\"at
              zu K\"oln, Cologne, Germany\\
              \email{saur@geo.uni-koeln.de}
         \and
         KTH Royal Institute of Technology, Stockholm, Sweden
         \and
         Laboratory for Atmospheric and Space Physics, University of Colorado, Boulder, USA
         \and
             Department of Earth and
              Planetary Sciences; Department of Physics and Astronomy,
              Johns Hopkins University, Baltimore, USA
         \and
         Institut f\"ur Astrophysik, Georg-August-Universit\"at, G\"ottingen, Germany
             }

   \date{\today}

 \abstract{}{}{}{}{} 
 
  \abstract
{Observations of auroral emissions are powerful means to remotely sense the space
  plasma environment around planetary bodies and ultracool
  dwarfs. Therefore successful searches and
  characterization of aurorae outside the Solar System will open
  new avenues in the area of extrasolar space physics.} 
   {We aim to demonstrate that brown dwarfs are ideal objects to search for UV aurora
  outside the Solar System. We specifically search for UV aurora on
  the late-type T6.5 brown dwarf 2MASS J12373919+6526148 (in the following 2MASS J1237+6526).
   }
   {Introducing a  parameter referred to as {\it auroral
  power potential}, we derive scaling models for auroral powers for
  rotationally driven aurora applicable to a broad
  range of wavelengths. We also analyze Hubble Space Telescope observations obtained with the
  STIS camera at near-UV, far-UV, and Ly-$\alpha$ wavelengths of 2MASS J1237+6526.}
   {We show that brown dwarfs, due to their typically strong surface magnetic
  fields and fast rotation, can produce auroral UV powers on the order of
  10$^{19}$ watt or more.
Considering their negligible thermal UV
  emission, their potentially powerful auroral emissions make brown dwarfs
  ideal candidates for detecting extrasolar aurorae.
We find 
possible
emission from 
2MASS J1237+6526,
but cannot conclusively attribute it to the brown dwarf
due to low signal-to-noise values
in combination with
nonsystematic trends in the background fluxes.
The observations provide upper limits for the emission at
  various UV wavelength bands.
The upper limits for the emission
correspond to a UV luminosity of 
  $\sim$1 $\times$ 10$^{19}$ watt, which lies in the range of the
  theoretically expected values.}
  {The possible auroral emission from the dwarf could be produced by a
  close-in companion and/or magnetospheric transport processes. }
   \maketitle

%

\section{Introduction}
Brown dwarfs are objects in the mass range between low mass stars and planets (13-80 M$_J$ Jupiter
masses;
see, e.g.,
\citealt{kuma62,haya63}).
They possess very strong magnetic fields and are often fast
rotators,  which is expected to lead to large
rotationally dominated magnetospheres \citep{pine17}. However, very
little is known about
the plasma properties within
the dwarfs' magnetospheres and the
physical processes that energize the plasma.
In the Solar System strong 
constraints on the planets' magnetospheres stem from
remote sensing of their auroral emissions through the full
electromagnetic wavelengths range.  By auroral emission we refer to  nonthermal
emission caused by electron or ion impact excitation of atmospheric neutrals where
the electrons and ions originate external to the bodies.
The upper atmosphere of a planet in which the auroral emission is
excited acts as a screen, and thus makes magnetospheric processes
visible on the central body.

Aurorae outside of the Solar System have not been detected with
certainty. H-$\alpha$ and radio emission
from the ultracool dwarf
2MASS J18353790+3259545
\citep{lepi03,reid03,berg06,berg08},
here referred to as LSR
J1835+3259,
was interpreted by \citet{hall15} as auroral emission. However,
subsequent observations with the Hubble Space Telescope (HST) to confirm the
ultraviolet (UV) counterpart to the expected auroral emission revealed a UV
spectrum that does not resemble the auroral emission known from
Jupiter, the brightest auroral emitter in the Solar System, but 
instead
rather
resembles the emission of a late-type cool star
\citep{saur18}. No other independent confirmation of auroral emission
from LSR J1835+3259 as proposed
by \citet{hall15} exists.
\citet{kao16,kao18} detected
circularly polarized radio emission from a number of
T- and L-dwarfs. In analogy
to circularly polarized radio emission caused by auroral electron
beams in the Solar System, these authors attribute the radio emission
from the dwarfs to auroral electric current systems. However, further confirmation is
needed of whether  electron or ion beams stemming
from magnetosphere-ionosphere coupling processes in the brown dwarfs'
magnetospheres are indeed  the cause of the circularly polarized radio
emission
or if the emission is chromospheric.

We obtained observations with the HST/STIS camera at near-UV (NUV), far-UV (FUV),
and Ly-$\alpha$ wavelengths of the brown dwarf
2MASS J12373919+6526148
\citep{burg99a}, here referred to as 2MASS
J1237+6526.
For the interpretation of these
observations in the context of UV aurora outside the Solar System,
we first need to develop a theoretical framework and context  
before the presentation of the data and its analysis.
Therefore, we derive auroral luminosity estimates
 and demonstrate that brown dwarfs are ideal candidates to
search for ultraviolet aurora outside of the Solar System. On the one
hand, they possess strong kilogauss magnetic fields 
and are
 fast rotators  with periods of only a few hours.
 We  show that these conditions favor strong auroral emission. On the other hand, brown dwarfs
 are not massive enough to burn hydrogen 
similar to
stars
and only have weak internal energy sources from fusion of
deuterium and lithium. Therefore, they are too cold
to generate competing thermal UV emission (for a detailed discussion of brown dwarf
properties, see, e.g.,
\citealt{chab00,burr01,kirk05,reid13}).

Expressions for radio (or auroral) emission from planetary bodies and dwarfs
have been derived or applied by a number of authors. For example,
\citet{zark01} and \citet{zark07} developed a radio-magnetic Bode's
law for planets and exoplanets.
The focus of these studies lay on emission driven by the flow of
magnetized plasma past the bodies 
 but not on rotationally driven
plasma transport within the objects' magnetospheres. A series
of authors estimated auroral powers generated by
a companion for individual cases
\citep[e.g.,][]{zark07,saur13,hall15,turn18,fisc19,veda20,veda20a}
and for radio emission from ultracool dwarfs due to their fast
rotation \citep[e.g.,][]{nich12,turn17}.

The brown dwarf 2MASS J1237+6526, for which we present HST
observations,
is a dwarf of spectral type
T6.5 in the constellation Draco. It was discovered by
\citet{burg99a} and
is located at a distance of
10.42
$\pm$0.52 pc
\citep{vrba04}.
2MASS J1237+6526 possesses an
effective temperature T$_{eff}$ of
$\sim$830$_{-27}^{+31}$ K \citep{kao16}.
Its mass was 
estimated to be 41 $\pm$26 M$_J$ and its age is considered to be older than 3.4
Gyrs \citep{fili15,kao16}. \citet{burg00} discovered
anomalously hyperactive H-$\alpha$ emission at log(L$_{H-\alpha}$ /L$_{bol}$) =
- 4.3 from 2MASS J1237+6526. Such intensive emission is typically seen in early
M-dwarf stars, but is exceptional for late-type brown dwarfs
\citep{gizi00,schm15,pine17}.
The origin of this exceptionally bright emission is still
unclear. Subsequent observations of 2MASS J1237+6526 confirmed that the
H-$\alpha$ emission is persistent over at least 1.6 yr with only mild
variations within a factor of two, ruling out flaring as a cause
\citep{burg02}. \citet{lieb07} also discarded a youthful chromospheric
activity or a massive, possibly binary, companion as
the
origin of the
H-$\alpha$ emission. Search for variability within the J band rendered
conflicting results \citep{burg02,arti03}. A highly interesting idea
was raised by \citet{lieb07} who suggested that the emission could be
caused by a smaller companion with T$_{eff}$ < 500 K based on a
possible color excess seen in Spitzer IRAC photometry. Important
constraints on 2MASS J1237+6526 stem from radio observations
by \citet{kao16,kao18}. The radio emission includes a highly circularly
polarized and pulsed component, which was used to derive the dwarf’s
rotation period of 2.28 hr \citep{kao18}. The radio emission provides
evidence that 2MASS J1237+6526 possesses a very strong average surface magnetic
field of at least 2.9 kG. In analogy with the circularly polarized
radio emission from the aurorae of the Solar System planets,
\citet{kao16,kao18} hypothesize that the circularly polarized radio
emission from 2MASS J1237+6526 may be auroral emission. However, it remains
unclear whether this emission is really caused by auroral
electrons that are energized outside of the dwarf’s
atmosphere, chromosphere, or ionosphere 
(i.e., in the dwarf’s
magnetosphere).

In this paper we first
discuss properties of the plasma and
magnetic field environment around brown dwarfs, planets, and low mass
stars in a broader context, and derive scaling models
for their potential auroral luminosities (Section
\ref{s:scaling}).
Then we  present the analysis of HST observations of the
brown dwarf 2MASS J1237+6526 (Section \ref{s:obs}). Subsequently, we
discuss what the HST observations of 2MASS J1237+6526 might imply in the
context of its possible auroral emission (Section \ref{s:discussion}). 

\section{Auroral power potential}
\label{s:theory}
Aurora
is defined as electromagnetic emission caused by impacting electrons or ions on
atmospheric neutrals, where the impacting particles are produced outside
the objects' atmospheres (i.e., usually in their magnetospheres). In this section we
derive estimates of potential auroral powers based on the underlying energetics of magnetospheric
processes, which are the root cause of auroral emission. We show that brown dwarfs are ideal objects to search for UV aurorae
outside the Solar System.

\subsection{Auroral power: Overview}
Auroral processes require three components:
(1) a generator that produces electric current and
electromagnetic power, (2) an accelerator where the electromagnetic
energy is converted into particle acceleration, and (3) an atmosphere
where the energized particles precipitate into and excite auroral
emission 
\citep[e.g.,][]{mauk12}.

The generators known in the Solar System are powered in one of three different
ways:   (i) driven by plasma flow external to the
planet's magnetospheres,
(ii) driven by plasma transport within the planets' fast rotating magnetospheres, or (iii) driven by moons
within the planets' magnetospheres. 

The aurora of Earth falls into category (i) and possess a solar wind driven aurora, where
the energy ultimately comes from the solar wind flow exerting
forces on Earth's magnetosphere and driving reconnection at the
magnetopause \citep[e.g.,][]{mauk12}.
Similarly, Ganymede's aurora is mostly powered by the flow of plasma
from Jupiter's magnetosphere against Ganymede's mini-magnetosphere \citep[e.g.,][]{evia01}.

Jupiter's main auroral emission is
a prime example of 
category 
(ii),   an aurora driven by internal magnetospheric plasma transport. In
the     case of Jupiter, its close-in moon Io is the root cause of
plasma production at a rate of about 1 ton/s
\citep[e.g.,][]{broa79,hill79,dess80,smit85}.
Jupiter's fast rotation 
(with
a period of about 10 hr) generates
strong centrifugal forces within its magnetosphere, which cause the
magnetospheric plasma to move radially
outward. Due to conservation of angular momentum the corotation of the
outward moving plasma breaks down with distance, which means that  the
magnetospheric plasma
does not fully corotate with Jupiter any more. The sub-corotating
plasma thus generates magnetic stresses that couple
Jupiter's magnetosphere to its ionosphere. This coupling drives angular momentum
and energy transport
between the two regions, which spins up the magnetosphere toward 
corotation, but slows down the angular velocity of its ionosphere.
In the case of Jupiter this process is the main auroral power
generator
\citep{hill01} even though many details of the coupling and the
physics of the particle acceleration are currently being investigated
via Juno spacecraft measurements
\citep[e.g.,][]{mauk17,clar18,saur18a}.
The underlying mechanism of how a rotating and expanding magnetosphere couples to
its ionosphere is similar to magnetic braking where an expanding
stellar wind magnetically couples to its star and subsequently slows down the
rotation of the star.

Aurorae of category (iii) are caused by moons within planetary
magnetospheres. The moons  are obstacles to the
magnetospheric plasma and perturb the plasma flow and magnetic
field. This also causes magnetic stresses, which propagate along the
planet's magnetic field and carry high energy fluxes between the
moons and their host planets \citep{gold69,neub80,goer80,zark07,saur13}.

In the following we only consider  auroral processes (ii)
and (iii), those due to magnetospheric mass transport and a planetary
companion. The first process (i),   flow external to the
magnetosphere, is not expected to be relevant for
brown dwarfs, 
due to the expected very large sizes of brown dwarf magnetospheres.
The sizes of
their magnetospheres are not known from direct observations, nor are
the detailed properties of the interstellar medium (ISM) surrounding
the brown dwarfs.
To estimate the sizes of the dwarfs' magnetospheres,
we assume ISM properties similar to those near
the heliosphere
from \citet{mcco12} with a magnetic field strength of
$B_{ISM} = 3 \times 10^{-6}$ gauss, a proton density $n_{ISM}$ of 0.07
cm$^{-3}$, and a relative velocity between the heliosphere and the ISM
of $v_{ISM}$ = 23.2 km s$^{-1}$.
We assume
that the magnetic pressure of
the dwarf's magnetosphere is
balanced by the sum of the magnetic and ram pressure of the ISM at the
magnetopause. Under the assumption of a dipole field for the brown dwarf, this balance
leads to a
location of the magnetopause in units of the brown dwarf's radius $R_{BD}$ given by
\begin{eqnarray}
\frac{r}{R_{BD}} =
\left(
\left(
\frac{B_{ISM}}{B_{BD}}
\right)^2
+
\left(
\frac{m_p \; n_{ISM} \; {v_{ISM}^2}}
{B_{BD}^2/2 \mu_0}
\right)
\right)^{- 1/6}
,\end{eqnarray}
with $m_p$ the
proton mass, $\mu_0$ the permeability of free space, and $B_{BD}$ the
equatorial field strength of the brown dwarf.
This
expression yields a magnetopause distance of
740 R$_{BD}$ for 2MASS J1237+6526.
Any flow of plasma inside the magnetosphere will shift the location of
the magnetopause farther away. 
The estimated gigantic size of the dwarf's
magnetosphere  can be assumed to be typical for strongly magnetized
brown dwarfs. The size implies that the 
interaction of the interstellar medium with 
this magnetosphere and similar dwarf magnetospheres 
occur at such large distances 
that auroral coupling  processes to the dwarfs' atmospheres are likely ineffective.

\subsection{Auroral power: Scaling models}
In this subsection we  derive scaling models for luminosities 
of aurorae driven by
magnetospheric mass transport or a planetary companion. These scaling
models are applicable to the various wavelength ranges employed in the
search and characterization of exo-aurorae. We   show
that it is possible to introduce a universal quantity, referred 
to as {\it auroral power potential} $S_{pot}$, which characterizes the
ability of an object to generate aurora.
The subsequent models assume for mathematical simplicity a pure
dipole magnetic field for the primary bodies, even though the planets in the Solar
System and low mass stars are known to possess magnetic fields with higher-order moments
\citep[e.g.,][]{conn18,mori08a,yada15,berd17}. Magnetic field
contributions from higher-order moments fall off more rapidly with distance than dipole
components, thus at the magnetospheric locations of the auroral generators the dipole
components likely dominate the local magnetic fields.
In this work the atmospheres of 
brown dwarfs are assumed to possess electrically conductive layers
similar to the atmospheres of the planets in the Solar System
\citep[e.g.,][]{rees89}.
Such ionized layers on brown dwarfs can form via the
interstellar radiation field or ionizing auroral electrons beams, for example,  and
have also been referred to as  ionospheres on brown dwarfs 
\citep{hell19}. These ionized layers provide free electrons and ions 
that can drive magnetospheric current systems and/or chromospheric heating
\citep{rodr18}.

\subsubsection{Aurora due to magnetospheric mass transport}
\citet{hill01} derived an expression, resulting from the mass transport in Jupiter's fast rotating magnetosphere,
 which characterizes
the energy flux between the magnetosphere and its ionosphere.
This energy flux is given by expression (4) in \citet{hill01} and
reads
\begin{eqnarray}
P_{mag,J} = 2 \pi \Sigma_J B_J^2 \Omega_J^2 R_J^4/{\hat{L}}_J^2
\label{e:Hill}
,\end{eqnarray}
with
Jupiter's dipolar equatorial surface magnetic field $B_J$, its angular velocity
$\Omega_{J}$, the ionospheric conductance
$\Sigma_{J}$, Jupiter's radius $R_{J}$, and the distance $\hat{L}_J$
between Jupiter and the magnetospheric region where the corotation
breaks down. The $\hat{L}$ parameter is
dimensionless and describes the breakdown distance in units of
$R_J$. In Jupiter's magnetosphere the breakdown occurs  around $\hat{L}_J$ $\sim$ 30 \citep{hill01}.
At the distance $\hat{L}_{J}$, the  energy
flux between the magnetosphere and the ionosphere maximizes.
We 
generalize expression (\ref{e:Hill}) to an arbitrary host body
with similar magnetospheric transport processes and rewrite the expression as
\begin{eqnarray}
P_{mag} = S_{pot} \left( \pi \frac{ R_{host}^2}{\hat{L}_{host}^2} \right) \Sigma_{host}
\label{e:Pmag}
,\end{eqnarray}
where the variables with the subscript ``$host$'' refer to the auroral host object
under consideration. The host can be a planet, a brown dwarf, or a
star. In 
expression 
 (\ref{e:Pmag}) we combined the quantities polar magnetic
field strength $B_{host}$, angular velocity $\Omega_{host}$, and radius $R_{host}$ of the auroral emitting host
to a new quantity, which we refer to as {auroral power potential:}
\begin{eqnarray}
S_{pot} = B_{host}^2 \Omega_{host}^2 R_{host}^2\;.
\label{e:S_pot}
\end{eqnarray}
This quantity is universal for the cases studied here,
meaning that it characterizes auroral power generated by magnetospheric mass transport
and by a companion, as shown in  this Subsection and in Section
\ref{ss:companion}, respectively.
The auroral power potential can
be written as $B_{host}^2 v_{host}^2$ with the velocity given by $v_{host} = \Omega_{host}
R_{host}$. Assuming the frozen-in field theorem ${\bf E} = - {\bf v} \times
{\bf B} $ with $\bf E$ the motional electric field in the nonrotating rest frame,
the auroral power potential
is simply $E_{host}^2$ and therefore proportional to the work per volume
element ${\bf j_{host}}\cdot
{\bf E_{host}} = \sigma_{host} E_{host}^2$ with  the conductivity 
$\sigma_{host}$ of the host's ionosphere.
Thus, the auroral
power potential is the power that can be exerted by the flow $v_{host}$  in
a magnetic field $B_{host}$ per unit conductivity and unit volume on the
surface of the host. The total power in (\ref{e:Pmag}) is then given
by integration of the power potential over the volume on the host
where the auroral coupling and emission occurs multiplied by its
conductivity. The conductivity $\sigma$ integrated along the magnetic
field lines, which are essentially radial
within the planet's polar ionosphere, leads to the conductances
$\Sigma_{host}$. Thus, 
Expression
(\ref{e:Pmag}) can be rewritten as
\begin{eqnarray}
P_{mag} = S_{pot} \; \underbrace{\Sigma_{host} \; A_{auroral} \; \hat{v}^2}_\text{Q}
\label{e:S_pot_a}
,\end{eqnarray}
with the area $A_{auroral}$ on the planet from which the auroral emission
occurs. In this model it is located at colatitude $\Theta $ given
by $\sin \Theta = \hat{L}^{-1/2}$ \citep{hill01}. It would correspond
to a width of $\Delta \Theta = \frac{1}{4} \hat{L}^{-1/2}$ of the auroral ovals. In units of degrees the location of
Jupiter's auroral oval is thus at
10.5$^{\circ}$
colatitude
with a width of 2.6$^{\circ}$ (based on $\hat{L}
= 30)$. 
Since the location
of the oval is at $\Theta$, 
the velocity $v$ of the ionosphere is
not $R_{host} \Omega$, but reduced by a dimensionless quantity,
$\hat{v} = v(\Theta)/(R_{host} \Omega_{host}) = \sin \Theta$.
We combine the last three terms in Eq.  (\ref{e:S_pot_a})  to a new quantity
Q, which represents the individual characteristics of the auroral
driver within a host's magnetosphere.
Thus, in simple words, the total power from which the aurora at various
wavelengths draws its power is given by the auroral power potential
$S_{pot}$ times a factor $Q$ which depends on the details of each
magnetosphere under consideration. The Q-factor is expected to vary
individually
and can be
different for magnetospheres with even the same power potential.
The  Q-factor 
depends, for example, on the plasma sources and their composition, the mass transport
in the magnetosphere, and the ionization state of the hosts.

It is instructive to compare the electromagnetic energy fluxes derived
here assuming the ionosphere of the host can be characterized by a
spinning disk with angular velocity $\Omega_{host}$, radius
$R_{host}$, and conductivity $\sigma$ within a magnetic field 
$B_{host}$ parallel to the spin axis of the disk. The work done per
unit time 
${\bf j} \cdot {\bf E}$ within the whole conducting disk is given by
\begin{eqnarray}
P_{disk} = \frac{1}{2} \pi \;\Sigma_{host} \; R_{host}^2 \; S_{pot} \;.
\label{e:disk}
\end{eqnarray}
The energy flux in Eq. (\ref{e:S_pot_a}) can be written in units of the
work done by a disk as
\begin{eqnarray}
P_{mag} = 2 \;\hat{A}_{auroral}\; \hat{v}^2 \; P_{disk}
\label{e:pmag_disk}
,\end{eqnarray}
with $\hat{A}_{auroral}$ the size of the auroral area in each hemisphere normalized to
 the size of the disk and $\hat{v}$ the normalized rotation velocity
 of the disk at the location of the auroral area.

The expressions Eq.  (\ref{e:Pmag}), (\ref{e:S_pot_a}), or
(\ref{e:pmag_disk}) 
describe in various ways the total energy flux between the host and
its magnetosphere due to radial mass transport. Only
a fraction of this energy flux is converted into
particle acceleration and subsequently into auroral emission at a
certain wavelength or wavelength band. We
characterize this fraction by an efficiency factor $\epsilon$. 
In the case
of Jupiter the total energy flux between its ionosphere and magnetosphere has been
estimated  as $P_{mag} = 3.1 \times$
10$^{14}$ watt by \citet{hill01}. Jupiter's auroral energy fluxes
within different wavelength bands and the associated fraction
$\epsilon_J$  are given
in Table \ref{table:epsilon}.
\begin{table*}
      \caption[]{Auroral luminosities at Jupiter and efficiency factor for
      various wavelength bands.
}
\label{table:epsilon}      
\centering          
\begin{tabular}{l r r r }     
\hline\hline
 wavelength & Luminosity $L_{mag}$ & Luminosity $L_{comp}$&efficiency $\epsilon$ \\
 & main oval [watt] & Io footprint [watt]  &\\
\hline
 X-ray &   1 -- 4 $\times$ 10$^9$ (a) & &0.3 --1.3 $\times$ 10$^{-5}$\\
Far-UV  &    2 -- 10 $\times$ 10$^{12}$ (a)   &0.4 -- 30 $\times$
10$^{10}$ (c) &0.6 -- 3 $\times$ 10$^{-2}$ \\
Near-UV &  2 -- 10 $\times$ 10$^{11}$ (a) & $\sim$5 $\times$10$^9$ (a)
& 0.6 -- 3 $\times$ 10$^{-3}$ \\
Visible &  1 -- 10 $\times$ 10$^{10}$ (a) &$\sim$3 $\times$ 10$^{8}$ (a) & 0.3 -- 3 $\times$ 10$^{-4}$ \\
IR &$\sim$50 $\times$ 10$^{12}$ (a)&3 -- 10 $\times$
10$^{10}$ (a) &$\sim$10$^{-1}$ \\
Radio & 1 -- 10 $\times$ 10$^{10}$(a,b)&1 -- 100 $\times$ 10$^{8}$ (a,d)&  0.3 -- 3 $\times$ 10$^{-4}$\\
\hline                  
\end{tabular}
\tablefoot{(a): \cite{bhar00}, (b): \cite{zark98}, (c) \cite{wann10},
  (d) \cite{zark07}. The X-ray scaling needs to be taken with
caution as the origin of
Jupiter's X-ray emission is not fully understood.
}
\end{table*}
We thus can write the auroral luminosity for a certain wavelength
range as  
\begin{eqnarray}
L_{mag} = S_{pot} \left( \pi \frac{R_{host}^2}{\hat{L}_{host}^2} \right)
\Sigma_{host} \epsilon_{host} \;.
\label{e:Lmag}
\end{eqnarray}
Alternatively, it is useful to write the auroral luminosity in Eq. (\ref{e:Lmag})
as a scaling model with respect to observed auroral luminosities of the
Jupiter system $L_{mag, J}$ within
certain wavelength bands as
\begin{eqnarray}
L_{mag }
=
\underbrace{
\left(
\frac{S_{pot,\;host}}{S_{pot,J}}
\right)
}_\text{$S_{pot,rel}$}
\underbrace{
\left(
\frac{\Sigma_{host}}{\Sigma_J}
\frac{\hat{L}_{host}^{-2}}{\hat{L}_{J}^{-2}}
\frac{R_{host}^2 }{R_J^2}
\frac{\epsilon_{host} }{\epsilon_J}
\right)
}_\text{$Q_{rel}$}
L_{mag, J}
\label{e:P_mag}
,\end{eqnarray}
where the subscript $J$ refers to values for Jupiter with S$_{pot,J}$ = 600 watt m$^{-2}$
siemens$^{-1}$
based on the values
from Table \ref{t:objects}. Values for Jupiter's ionospheric
conductances $\Sigma_J$ from the literature lie around 1 
siemens
(e.g., 0.6 siemens in \citealt{hill01} or 5 siemens in \citealt{stro83} and
$\hat{L}_J$ = 30 \citealt{hill01}).
%
Note that in this work $\hat{L}$ refers to the normalized location of
corotation break down while other variables $L$ without ''hat'' refer
to the various luminosity . 
 The first term on the right-hand side
describes the power potential relative to Jupiter $S_{pot,rel}$.
The other quantities within the second set of parentheses combined as
$Q_{rel}$ in Eq. (\ref{e:P_mag}) describe auroral generator characteristics
relative to Jupiter. 
$S_{pot}$ and $Q_{rel}$ are independent, which means that  a host
can have in principle a larger auroral power potential, but in the
absence of a magnetospheric plasma source Q can still be zero.
The Q factor includes the location where the corotation breaks down 
(${\hat{L}}_{host}$),
 the ionospheric conductance ($\Sigma_{host}$, i.e., the
altitude integrated conductivity $\sigma$), and the
efficiency ($\epsilon_{host}$), which all depend on the details of the mass sources
within the magnetosphere and the ionospheric properties of the host
system.  Possible ionization can come from external ionizing UV radiation or internal
impact ionization, which depend on the individual system
\citep{rodr18}. Aside from
the radius, which for brown dwarfs can be assumed to be around
Jupiter-size, the other parameters of Q are difficult to
estimate without further information. Observations of auroral emission,
however, will allow us to constrain them.

\subsubsection{Aurora due to a planetary companion}
\label{ss:companion}
In case of a planetary companion within the host's
magnetosphere, the associated auroral processes can be
described by two different models, depending on the plasma density in
the magnetosphere. For very low densities and nearly unpopulated magnetospheres,
the coupling between the planetary companion and the central body
is given by the unipolar inductor model \citep{gold69};  for higher
densities the coupling turns into the Alfv\'en wing model
\citep{neub80,goer80,saur04}. The conditions for the transitions
between the two models are discussed in detail in
\citet{neub98}. It is unclear whether magnetospheres of brown dwarfs
are filled with plasma or are mostly empty. The calculations below show
that a companion in either case will generate an electromagnetic
coupling and auroral phenomena on the host body.

The energy flux in the unipolar inductor model was   derived by
\cite{gold69} as
\begin{eqnarray}
P_{uni} = \pi \Sigma_J E_J^2 (X+Y)Y
\label{e:uni}
,\end{eqnarray}
with the motional electric field $E_J$ and X and Y the semimajor and
semiminor axes of the resulting electromagnetic flux tube originating
between Io and Jupiter with the values taken at the surface of Jupiter. We can
rewrite Eq. (\ref{e:uni}) and generalize it  to an arbitrary companion
orbiting a host. We also introduce an efficiency factor $\epsilon_{host}$ for converting the energy
flux into auroral emission within a specific wavelength band. This
leads to
\begin{eqnarray}
L_{uni}  = S_{pot}
\left(
\frac{3 \pi }{4}  \frac{R_{comp}^2}{\hat{a}_{comp}^4}
\Sigma_{host} \epsilon_{host}
\right)
\label{e:Luni}
,\end{eqnarray}
with  the radius $R_{comp}$ of the companion
and the distance $\hat{a}_{comp}$ between the planet and the companion in
units of the radius of the host body $R_{host}$. The values for $\epsilon$ for the auroral imprint of
Io in Jupiter's atmosphere are similar to those of Jupiter's main
aurora \citep[based on][]{bhar00}.

The expression for the energy flux within the
Alfv\'en wing model was derived by \cite{saur13} as
\begin{eqnarray}
P_{Alf} =  2 \pi R_{Io}^2 \frac{E_{Io} B_{Io}} {\mu_0} M_A
\label{e:Alf}
,\end{eqnarray}
with
the motional electric field $E_{Io}$ and Jupiter's magnetospheric field
strength $B_{Io }$, both at the location of Io,
the radius $R_{Io}$ of Io,  
the permeability of free space $\mu_0$  and the Alfv\'en Mach number
$M_A$, i.e., the ratio of the flow
velocity $v$ to the Alfv\'en velocity $v_A$ in Jupiter's magnetospheric
plasma at the location of Io. In expressions (\ref{e:uni}) and (\ref{e:Alf}) it is assumed
that Io possesses a highly electrically conductive
ionosphere. We again can generalize the expression (\ref{e:Alf}) to an
arbitrary companion and host and calculate the resultant luminosity
$L_{Alf}$
\begin{eqnarray}
L_{Alf} = S_{spot}
\left(\frac{\pi}{2}  \frac{R_{comp}^2}{\hat{a}_{comp}^4}
\right)
\Sigma_{A} \epsilon_{host}
\label{e:LAlf}
,\end{eqnarray}
with an auroral efficiency factor
$\epsilon_{host}$ and Alfv\'en conductance $\Sigma_A = 1/(
\mu_0 v_A)= \sqrt(\rho
\mu_0)/B $, with   the magnetospheric field
strength $B$, and the mass density $\rho$ at the location of the companion.
 In the case of Jupiter, $\Sigma_A$ and $\Sigma_{host}$
assume comparable values  \citep{stro83}.
In this derivation  for simplicity we neglect the angular velocity of the companion compared
to the usually significantly higher angular velocity of the fast
rotating planets or brown dwarfs. This however could easily be fixed
by replacing $\Omega$ with the relative angular velocity
between the rotation of the host and the angular velocity of the
companion. The radius of the companion is its
effective radius (i.e., the radius including the companion's atmosphere
and ionosphere). When the companion possesses an internal magnetic
field, the effective radius can be significantly enhanced
\citep[e.g.,][]{neub98,saur13}, up to $\sqrt{3} R_M$ with
$R_M$ being the radial distance of the companion's magnetopause. In
this case, reconnection at the companions magnetopause generates
electrical conductivity of the object.

The energy fluxes induced by a companion
in the unipolar inductor case (\ref{e:uni}) and Alfv\'en wing case
(\ref{e:Alf}) can jointly be written as
\begin{eqnarray}
P_{comp} = \gamma \pi \; \Sigma R_{comp}^2 \;\hat{a}^{-4} \;S_{pot} \nonumber
\\
= 2 \gamma \; \Sigma \; \hat{v}^2 \;A_{eff} \;S_{pot} \nonumber \\
= 4 \gamma \; \hat{v}^2 \; \hat{A}_{eff} \; P_{disk}
\label{e:combined}
\end{eqnarray}
with $\gamma = 3/4$ for the unipolar inductor case and $\gamma = 1/2$
for the Alfv\'en wing case. The conductance $\Sigma$ may represent  either the conductance of
the host body's ionosphere $\Sigma_{host }$ in the unipolar inductor case
or the Alfv\'en conductance $\Sigma_A$. The parameter 
$A_{eff}$ is the effective area of the
companion mapped along the host's magnetic field line onto the
host, and  $\hat{A}_{eff}$ is  $A_{eff}$ normalized to $\pi
R_{host}^2$.
Expressions in (\ref{e:combined}) show why it
is better to characterize the ability of a host to exhibit aurora with
a local quantity such as $S_{pot}$ on the host rather than a global quantity such as
the power of a rotating disk $P_{disk}$. Hosts with different sizes but
the same power potential would lead to the incorrect impression that they will
produce different auroral emissions for companions
with the same properties.

The resulting luminosities for the unipolar inductor and the Alfv\'en
wing model can now also be generalized to those caused by an arbitrary companion within a 
rotating magnetosphere. Using
(\ref{e:combined}), we find for the auroral luminosities caused by a
companion in units of
the auroral 
luminosity $L_{Io}$ caused by Io
\begin{eqnarray}
L_{comp}
=
\underbrace{
\left(
\frac{S_{pot,\;host}}{S_{pot,J}}
\right)
}_\text{$S_{pot,rel}$}
\underbrace{
\left(
\frac{\Sigma}{\Sigma_J}
\frac{R_{comp}^2 }{R_{Io}^2}
\frac{a_{comp}^{-4}}{a_{Io}^{-4}}
\frac{\epsilon_{host} }{\epsilon_J}
\right)
}_\text{$Q_{rel}$}
L_{Io} \;.
\label{e:P_companion}
\end{eqnarray}
The luminosity due to a companion thus  can also be written as the
product of the power potential of the host relative to the one of
Jupiter $S_{pot,rel}$ and a relative $Q_{rel}$ factor considering the
properties of the companion. The auroral luminosities caused by Io in
Jupiter's atmosphere are displayed in Table
\ref{table:epsilon}. Io's induced emission depends
on its position in Jupiter's magnetosphere, which leads to the range
of parameters in the table. 
The efficiency factors for Io's
auroral footprint are similar to those of the main oval based on theoretically
expected fluxes of 300-2000 $\times 10^9$ watt from \citep{saur13} and therefore not
shown separately in Table \ref{table:epsilon}.

Expressions (\ref{e:P_mag}) and (\ref{e:P_companion}) show that
the power for both auroral processes (i.e., magnetospheric transport or
a companion) is proportional to the quantity
$S_{pot,host}$, which describes the potential for an object to exhibit
auroral emission. The 
terms 
in the second set of parentheses, the Q-factor,
contains the properties of the magnetospheric mass transport,
the size and the distance of the companion plus its local plasma
environment, and 
the efficiency factor $\epsilon$. 

In the next subsection we investigate the
effects of the auroral power potential relative to Jupiter for
objects where $\Omega$, $B$, and $R$ are known. This gives a
guideline for selecting target objects to detect auroral
emission.  Measurements of auroral fluxes from these
objects will be a powerful tool to constrain individual properties of
the system included in the parameter Q (Eqs. \ref{e:P_mag} and
\ref{e:P_companion}).

\subsection{Auroral power potential: Brown dwarfs, planets, and stars}
\label{s:scaling}
   \begin{figure*}
   \centering
   \includegraphics[width=14.cm]{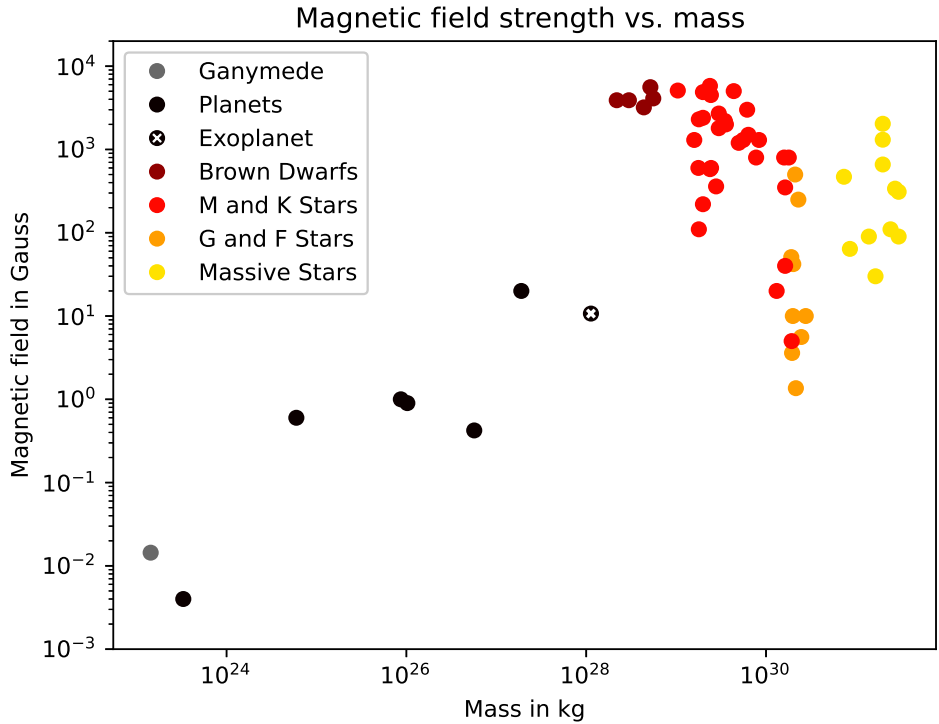}
   \caption{Observed surface magnetic field
   strengths or lower limits of surface fields (depending on
   observation method, see Appendix \ref{a:2}) as a function of mass for selected
   planetary bodies,
   brown dwarfs,
   and stars. The names of the objects with further details and
   associated references are provided in Appendix \ref{a:2}. 
   }
   \label{f:magnetic_fields}%
    \end{figure*}

   \begin{figure*}
   \centering
      \includegraphics[width=14cm]{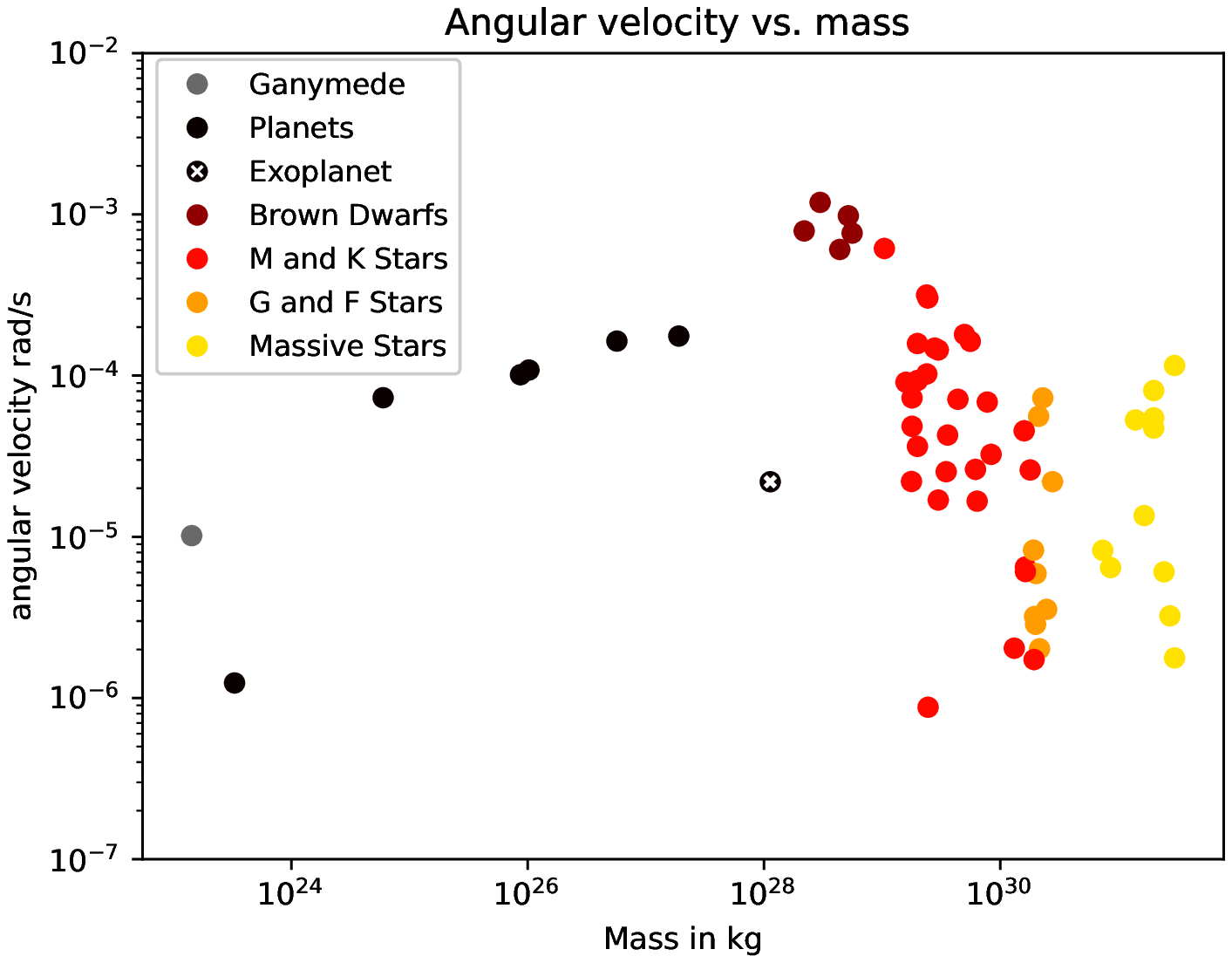}
      \caption{Angular velocity as a function of mass for selected
   magnetized planetary bodies, brown dwarfs,
   and stars. The names of the objects with further details and
   associated references are provided in Appendix \ref{a:2}.
   }
   \label{f:omega}%
    \end{figure*}
   \begin{figure*}
   \centering
      \includegraphics[width=14cm]{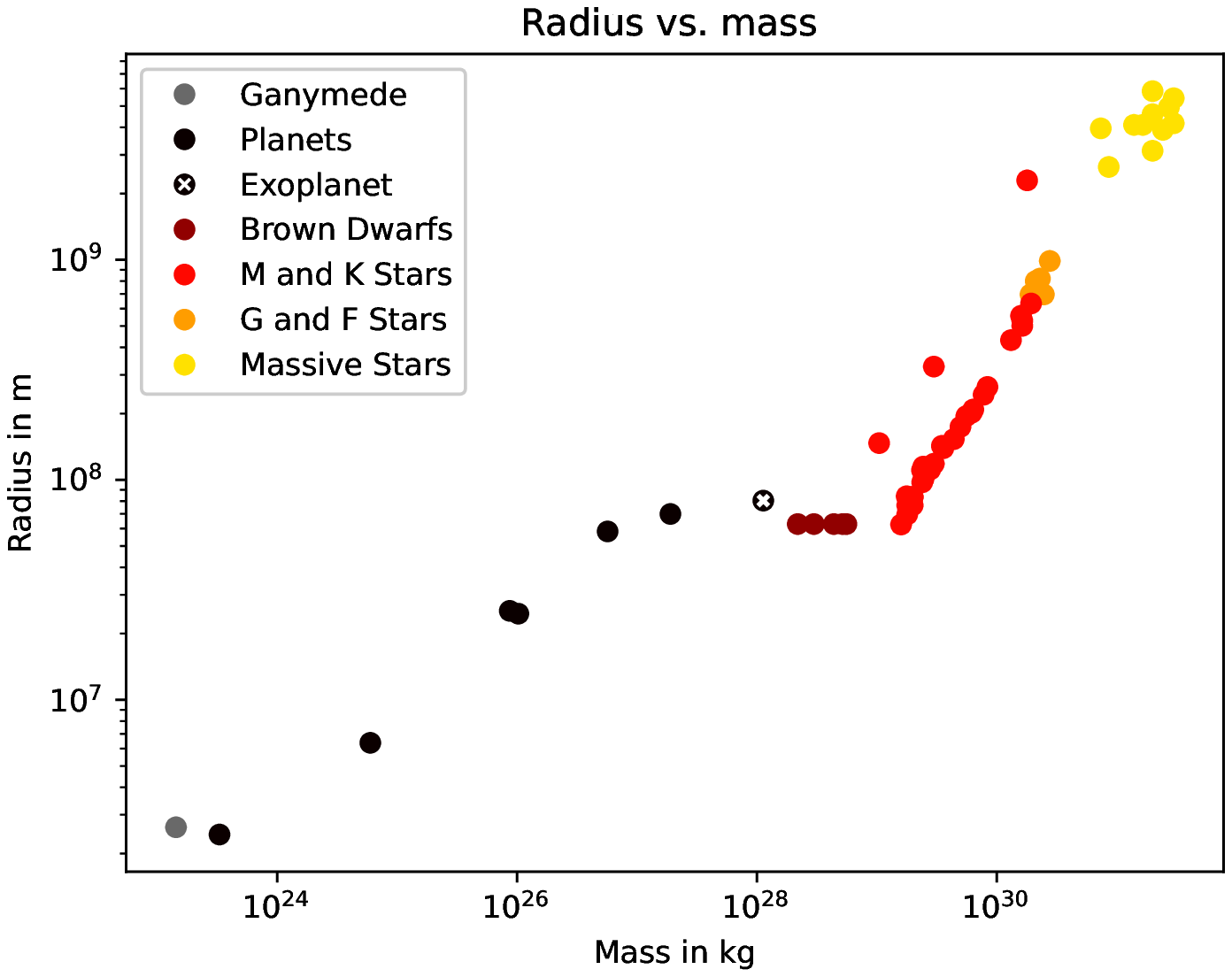}
      \caption{Radius as a function of mass for selected magnetized planetary bodies, brown dwarfs,
   and stars. The names of the objects with further details and
   associated references are provided in Appendix \ref{a:2}.
   }
   \label{f:radius}%
    \end{figure*}
   \begin{figure*}
   \centering
   \includegraphics[width=14cm]{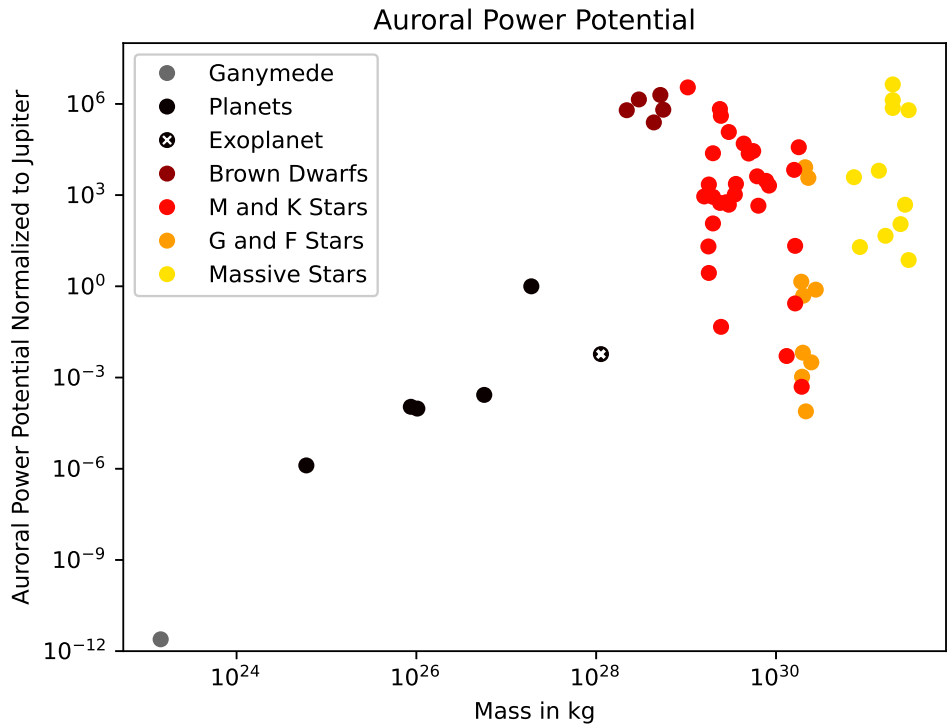}
   \caption{Auroral power potential normalized to the auroral power of
   Jupiter for aurora types that are driven by radial mass transport within strongly
   rotating magnetospheres  or by planetary companions (see text and
   Appendix \ref{a:2}).
   }
   \label{f:power_potential}%
    \end{figure*}
%
To investigate the role of the auroral power potential $S_{pot}$ we selected a
number of bodies covering the mass range of planets,
brown dwarfs, and stars (see Table \ref{t:objects}).
Significant auroral emission in the Solar System only occurs
on planetary bodies with intrinsic
magnetic fields. Intrinsic dynamo fields exist for Mercury, Earth, Jupiter,
Saturn, Uranus, and Neptune and for Jupiter's moon Ganymede.
Recently, a tentative detection of a magnetic field on the
   exoplanet {\ensuremath{\tau}} Bo{\"o}tis b based on radio
   observations with LOFAR by 
\cite{turn21} has been reported, which we also include.
For these
planetary bodies, together with brown dwarfs and low to intermediate
mass stars,
their
magnetic field strength as a function of the object's mass is displayed
in Figure \ref{f:magnetic_fields}. The magnetic fields have been
determined 
with 
various methods, such as  in situ measurements for the
Solar System targets, observations of radio emission near the electron cyclotron
frequency for brown dwarfs and {\ensuremath{\tau}} Bo{\"o}tis b,
or 
with methods based on the Zeeman-effect
for the stars
(see
last column in Table \ref{t:objects} for the explicit method applied for
each target). 
Some of these values might be lower limits  as  radio
emission, for example, often stems from magnetospheric layers above the objects' surfaces.
Despite these different techniques, Figure \ref{f:magnetic_fields}
still provides a general overview of the magnetic field strength for
 the various classes of objects. We include a wide spread of masses
   from small planetary objects to high mass stars for a basic
   visualization of the overall field strength distribution.
   The largest magnetic fields 
typically occur within the brown dwarf mass range into the low mass
   star regime, at least for the
cases where magnetic fields have been observed. Similarly the rotation
periods of brown dwarfs are typically very short,
on the order of hours
\citep{pine17,tann21},
which results in the high angular velocities in the
brown dwarf mass regime  displayed in Figure \ref{f:omega}. The
radii of the objects on average grows with mass, as displayed in
Figure \ref{f:radius}. However, in the brown dwarf mass regime the
radius-mass relation
is degenerated due to gravitational forces that alter
the balance between electrostatic attraction and
electron degeneracy pressure.
Thus, brown dwarf radii are
fairly constant
at values of about 0.9 $R_{J}$ \citep{vrba04,kao16,hatz15}.

With the magnetic field, the angular velocity, and the radius from each
body in Table \ref{t:objects}, we now calculate the
auroral power potential using expression (\ref{e:S_pot}) for the
selected bodies.  The results are displayed in Figure \ref{f:power_potential}.
The auroral power potential 
shows a maximum for brown dwarfs and very low mass stars.
It can be six orders of magnitude
larger than  Jupiter. Jupiter's auroral UV luminosity is in the
range of 2-10 $\times$ 10$^{12}$ watt \citep{bhar00}. Thus, brown dwarfs
have the potential for UV luminosities on the order of  10$^{19}$ watt.
The real auroral powers however can be
different,  higher or lower, as they additionally depend on the 
properties included in the $Q_{rel}$-factors in expressions
(\ref{e:P_mag}) and  (\ref{e:P_companion}).
B stars tend to have large auroral power potential values
similar to those of brown dwarfs due to their particularly large radii
(Fig \ref{f:radius}).

The auroral power potentials can be compared with the real auroral
powers of the planets and the moon Ganymede. The three lowest mass
objects are of terrestrial type:  the moon Ganymede, and the
planets Mercury and Earth.  Mercury does not
exhibit aurora because it does not possess an atmosphere. 
Ganymede's auroral emission has been observed in
the  FUV in the oxygen lines \ion{O}{I} 1304 and 1356 \AA$\;$
\citep[e.g.,][]{hall98,feld00,saur15}. Based on the measured fluxes,
we can derive a FUV luminosity of 3 $\times$ 10$^7$ watt, i.e., a factor
of 
$\sim$10$^5$ 
weaker than Jupiter's. Earth's auroral
luminosity is a factor of about 10$^3$ lower than that of Jupiter
\citep{bhar00}. However, the aurorae of these objects are driven
by reconnection and plasma flow outside of their respective magnetosphere,
and thus do not fall into the category  driven by internal mass flow
or driven by a companion. 

The outer planets Jupiter, Saturn, Uranus,
and Neptune are all fast rotators, possess magnetic fields, and
exhibit auroral emissions. Saturn's auroral UV power is 50 $\times$ 10$^9$
watt, that of Uranus  40 $\times$ 10$^9$
watt, Neptune's only 0.1 $\times$ 10$^9$
watt \citep{bhar00}. The auroral UV luminosities of Saturn and Uranus are
thus 10$^2$ to 10$^3$ times smaller compared to that of  Jupiter. This fits well into our
estimates shown in Figure \ref{f:power_potential} in particular given
that they are somewhat lower compared to Jupiter. The low UV
luminosity of Neptune, despite its auroral potential,  is generally
attributed to a very weak generator (i.e., very low mass transport
rates in the Triton torus) \citep{bhar00}.

\subsection{Thermal UV emission}
   \begin{figure*}
   \centering
   \includegraphics[width=14cm]{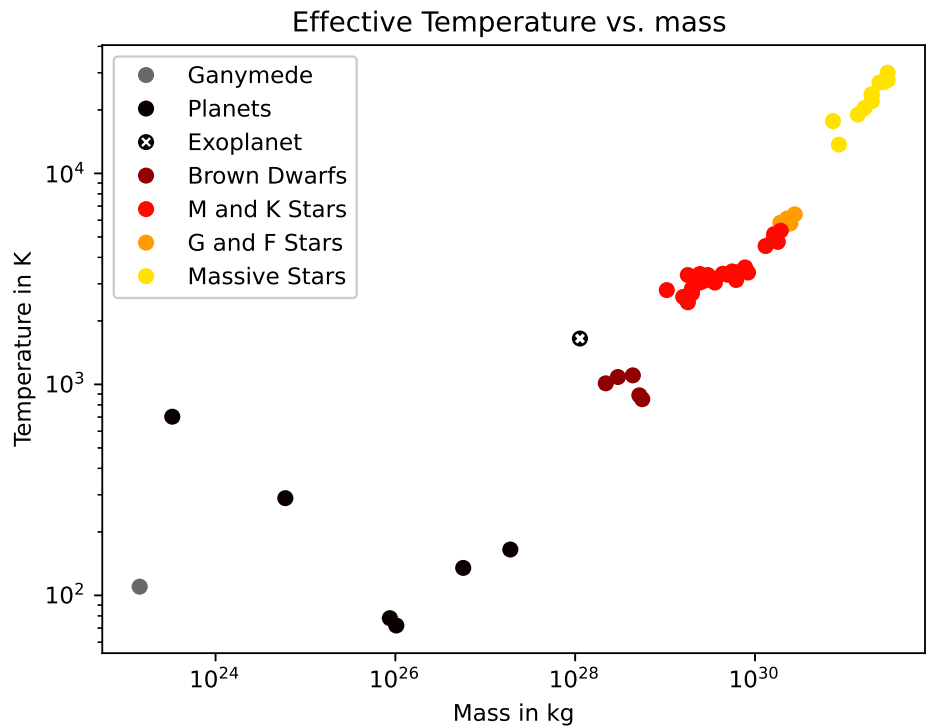}
   \caption{
Effective temperatures of selected magnetized objects (see table
\ref{t:objects}) as a function of mass.
   }
   \label{f:Teff}%
    \end{figure*}

   \begin{figure*}
   \hspace*{-1.cm}
   \includegraphics[width=18cm]{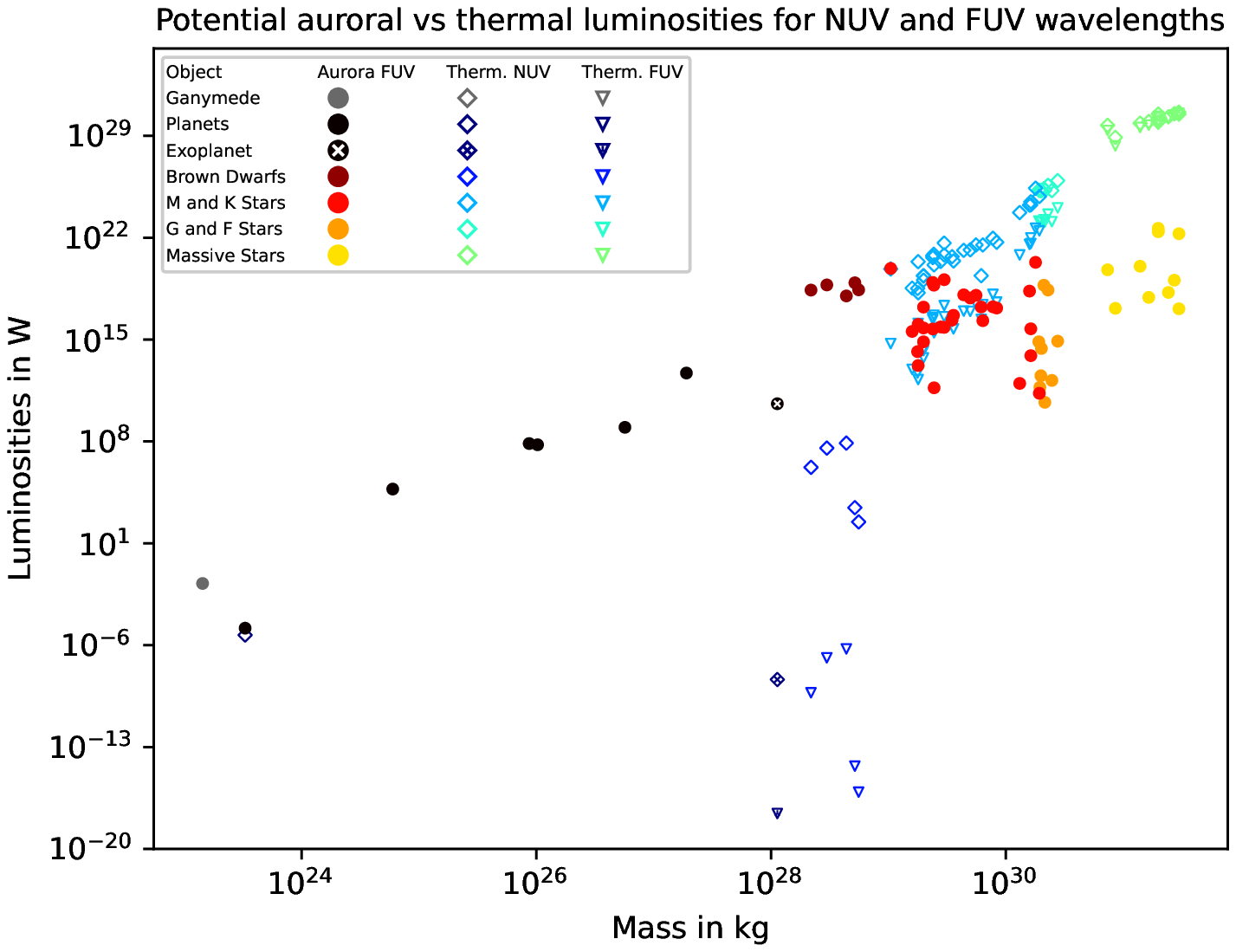}
   \caption{
   Thermal UV luminosities for the NUV wavelength range (empty diamonds)
   and FUV wavelength range
   (empty triangles) compared to expected auroral FUV emission (full
   circles).
   For Brown dwarfs, the massive outer planets of the solar
   system, the exoplanet {\ensuremath{\tau}} Bo{\"o}tis b
   and the moon Ganymede, the thermal UV emission is negligibly
   small compared to the possible auroral UV emission.
}
   \label{f:L_UV-compare}%
    \end{figure*}

Brown dwarfs are particularly well suited targets to search for UV
aurora outside the Solar System because, in addition to their large
auroral power potential, they exhibit low effective
temperatures.
In Figure \ref{f:Teff} we display the effective temperatures of the
objects spanning the mass range from planetary objects to massive
stars. The surface temperatures of Ganymede and the planets are primarily
controlled by solar irradiance and the thermal IR emission.
Effective temperatures of
Y- to L-type brown dwarfs lie in the range 
250
-- 2100 K, while
M dwarfs, which are usually low mass stars, lie in the range 2100 --
3500 K
\citep[e.g.,][]{fili15,luhm14}.
Solar-type stars possess effective
temperatures in the range of 5,000 K to nearly 10,000 K and massive
stars temperatures well above 10,000 K (references for each object are provided
in Table \ref{t:objects}). 

The expected thermal UV power strongly decreases with temperature of
stars and brown dwarfs, and thus also roughly decreases with mass.
The T6.5 dwarf 2MASS
J1237+6526 possesses a temperature of 830 K
\citep{kao16}.
Assuming only thermal emission
from the target we would expect a total luminosity of 10$^{2}$ watt in the  NUV within
1570-3180 \AA, and a total luminosity of 
10$^{-16}$ W
in the  FUV within 1150-1730 \AA. Both energy fluxes are entirely
negligible compared to the potential auroral UV emission of
10$^{19}$ W from the dwarf.

To further illustrate the competing effects of thermal UV emission
compared to possible auroral emission, we display in Figure
\ref{f:L_UV-compare} the thermal FUV and thermal NUV
for each object
assuming a blackbody emission with the effective temperatures from 
Figure \ref{f:Teff}. Due to the highly nonlinear temperature dependence
in Planck's law the thermal luminosity varies over more than 40 orders
of magnitudes from the brown dwarf to the massive star regime.
The thermal UV emission from the outer planets and Ganymede is so
small that it is below the displayed range. Only Mercury, due to its
proximity to the sun, is within the plot range of Figure \ref{f:L_UV-compare}.
In comparison to the thermal UV luminosities,
we show possible FUV luminosities
applying expression (\ref{e:P_mag}) with $Q_{rel} = 1$ and the average
FUV luminosity of Jupiter $L_{mag,J}$ from Table \ref{table:epsilon}. We do not show
the NUV auroral luminosity because it is  a constant factor of 10
lower than the FUV and would only overcrowd the figure.

Figure \ref{f:L_UV-compare} clearly shows that for the outer planets
and brown dwarfs the thermal UV is orders of magnitude below possible
auroral UV luminosities.
For early-type M-dwarf stars and more massive K-type stars the thermal UV emission 
starts to become a competing flux. G- and F-type stars possess thermal UV
emission typically in excess of possibly expected UV auroral emission.
This implies that auroral effects on a
solar-type  star will not be detectable within the NUV and FUV
bands. However, if a significant fraction of the electromagnetic
energy fluxes caused by the coupling of a companion is
concentrated in line emission from its star, it might still be possible. Such
observations do indeed exist. For example, \citet{shko05,shko08}
observe an excess in \ion{Ca}{II} emission on HD 179949, {\ensuremath{\tau}} Bo{\"o}tis,
and {\ensuremath{\nu}} Andromedae in synchrony with the orbital period of their close-in
planetary companions. The energy fluxes associated with the
\ion{Ca}{II} surplus is estimated to be  on the order of
$\sim$10$^{20}$ W. This phenomenon is referred to as
star--planet interaction (SPI)
\cite[e.g.,][]{lanz08,stru15,fisc19}. The energy flux models for the
various scenarios in this work are applicable to the
SPI studies.
For massive stars, possible auroral UV luminosities are
about 10 orders of magnitude lower than their thermal UV
luminosities, and thus are expected to be undetectable.
We note
 that the description of the thermal emission from one layer with
a single effective temperature is a first-order approximation only, but
 should still provide a reasonable estimate 
for the
detectability of auroral UV emission from outside the Solar System
against the object's thermal UV emission.

\section{HST Observations of brown dwarf 2MASS 1237+6526}
\label{s:obs}

After providing a basic framework for possible auroral emission from brown
dwarfs in Section \ref{s:theory}, we now analyze HST
observations of the brown dwarf 2MASS J1237+6526 \citep{burg99a}.
The dwarf  was observed with HST/STIS in 2020 during
two HST visits, with two orbits in visit 1 and three orbits in visit 2,
in order to search for UV auroral emission. The exposure details of
these observations are given in Table \ref{table:1}. One orbit was
dedicated to observation at NUV wavelengths, while Ly-$\alpha$ and FUV
wavelengths were equally split up within four orbits in order to
search for time variations in the aurora.
\begin{table*}
      \caption[]{Exposure details of HST/STIS observations of brown
   dwarf 2MASSS J1237+6526 (ID: 15870).}
\label{table:1}      
\centering          
\begin{tabular}{c c c r r r r l  l l}     
\hline\hline       
Visit & Orbit & Exp & Root name &UT Obs. Date & UT Obs. Time$^a$ & Exp. Time 
&Type &Grating & Disp.\\ 
\hline
\# & \# & \# & &yyyy-mm-dd & hh-mm-ss & s & & & \AA$\;$ pxl$^{-1}$ \\
\hline                    
   1 & 1 & 1 & oe1g01010 & 2020-01-16 & 19:54:07 & 814 & Ly-$\alpha$&G140M & 0.053 \\  
   1 & 1 & 2 & oe1g01020 & 2020-01-16 & 20:11:09 & 807 &Ly-$\alpha$ &G140M & 0.053\\
   1 & 2 & 3 & oe1g01030 & 2020-01-16 & 20:32:52 & 1475 & NUV       &G230L & 1.584\\
   1 & 2 & 4 & oe1g01040 & 2020-03-10 & 21:24:19 & 1475 & NUV       &G230L & 1.584\\
   2 & 1 & 5 & oe1g02010 & 2020-03-10 & 09:37:12 & 872 & FUV        &G140L & 0.584\\
   2 & 1 & 6 & oe1g02020 & 2020-03-10 & 09:55:10 & 871 & FUV        &G140L & 0.584\\               
   2 & 2 & 7 & oe1g02030 & 2020-03-10 & 10:13:07 & 1404& FUV        &G140L & 0.584\\
   2 & 2 & 8 & oe1g02040 & 2020-03-10 & 11:12:39 & 1055& Ly-$\alpha$&G140M & 0.053\\       
   2 & 3 & 9 & oe1g02050 & 2020-03-10 & 11:38:30 & 1404& FUV        &G140L & 0.584\\
   2 & 3 & 10& oe1g02060 & 2020-03-10 & 12:48:04 & 1159&Ly-$\alpha$&G140M  & 0.053\\      
\hline                  
\end{tabular}
\tablefoot{
$^a$: At the beginning of the exposures.
}
\end{table*}

The results of our analysis are presented for each of the three
wavelengths ranges in the next three subsections.
The detailed steps of the data analysis
are similar to those of the ultracool dwarf
LSR J1835+3259
presented in \citet{saur18} and described in Appendix
\ref{a:data_analysis}.
The target is extremely faint in the UV and cannot be located on the detector by
simple visual inspection. Therefore, we worked with the reference
positions provided in the header of the data files (see
Appendix \ref{a:data_analysis}).
Details of the acquisition of the faint target are 
provided in Appendix \ref{a:acq}.
The error analysis and the associated calculation of the
signal-to-noise (S/N) 
is based
on  the variance within the background fluxes in the
x2d-files, which provide the data at the highest calibration level (Appendix \ref{a:data_analysis}).
The resultant standard deviation $\sigma$ is then compared
with the extracted signal from the target. We
chose to work with the variance in the background because it covers
nonsystematic,
 nonphysical contributions in the background, due to the standard
processing pipeline performed by STScI, which can lead to large trends
in the background fluxes or artificially negative
values (see, e.g., \citet{saur18} or the STIS instrument handbook \citep{rile17}).
The error analysis is performed in parallel  based on the counts in the
flt-files, which provides the data at a lower calibration level (Appendix \ref{a:data_analysis}).
The two methods are not expected to give identical results due to
the processing steps of the STScI pipeline converting flt to x2d files,
for example  due to the corrections in remapping resolutions
elements between the flt and x2d data. The variance method
also underestimates the statistical uncertainty in the case of large background
fluxes as the method does not take into consideration the statistical
uncertainty of the variance. Therefore it is worthwhile to provide
S/N values from both methods.

\subsection{Near-UV spectrum}
%
   \begin{figure*}
   \centering
   \includegraphics[width=15cm]{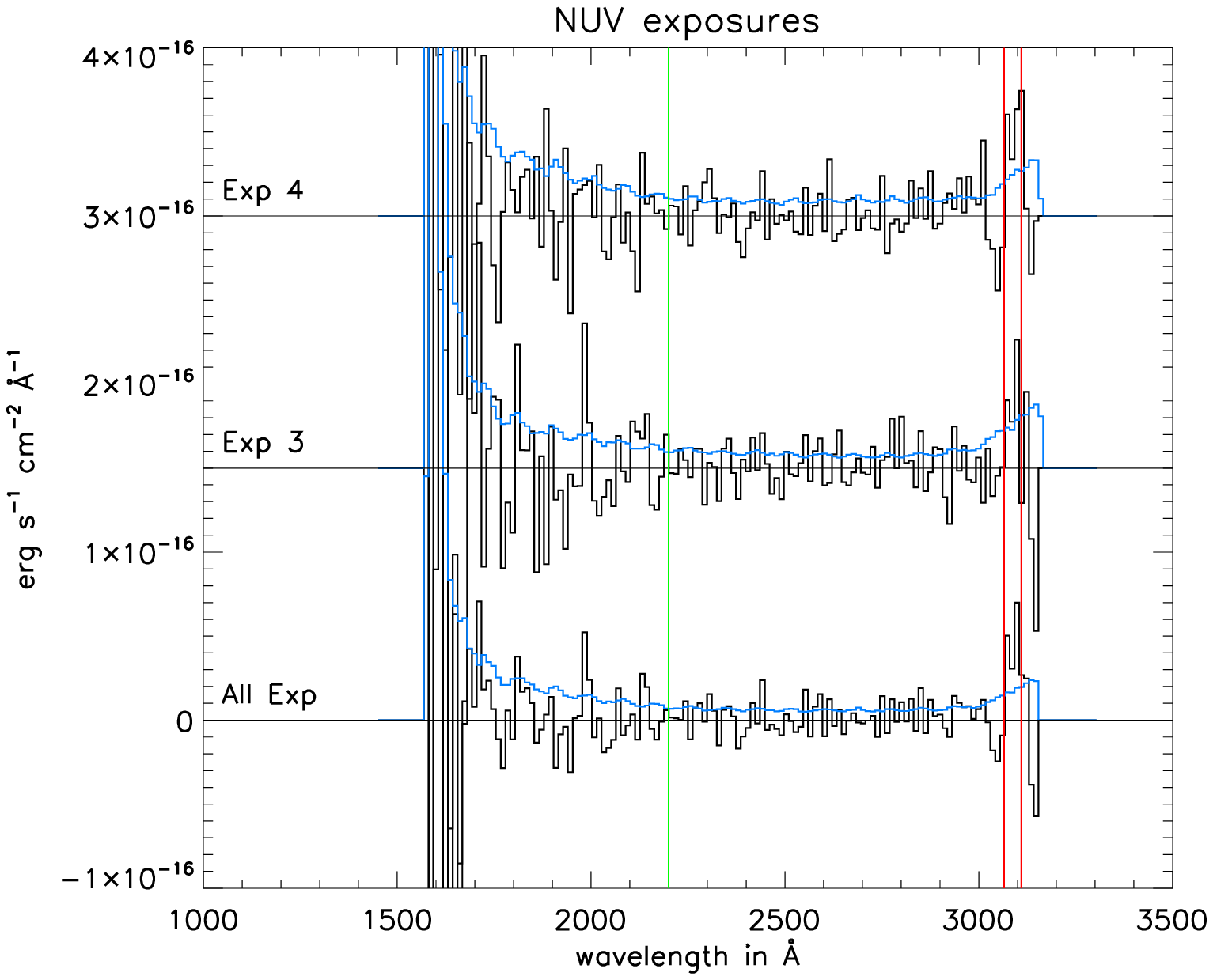}
   \caption{Near-UV spectra of exposures 3 and 4, and  combined
   spectrum. The spectra are separated by constant offsets for display
   purposes. The two red vertical lines indicate the wavelength
   band 3065 - 3110 \AA$\;$ displaying the most significant flux within the
   spectra, which might stem from TiO emission. The green line
   indicates 2200  \AA$\;$ (see text). In light blue we display
   the uncertainty spectrum calculated with the variance method described in
   Appendix \ref{a:data_analysis}.}
   \label{f:NUV_Spectrum}%
    \end{figure*}
   \begin{figure*}
   \centering
   \includegraphics[width=15cm]{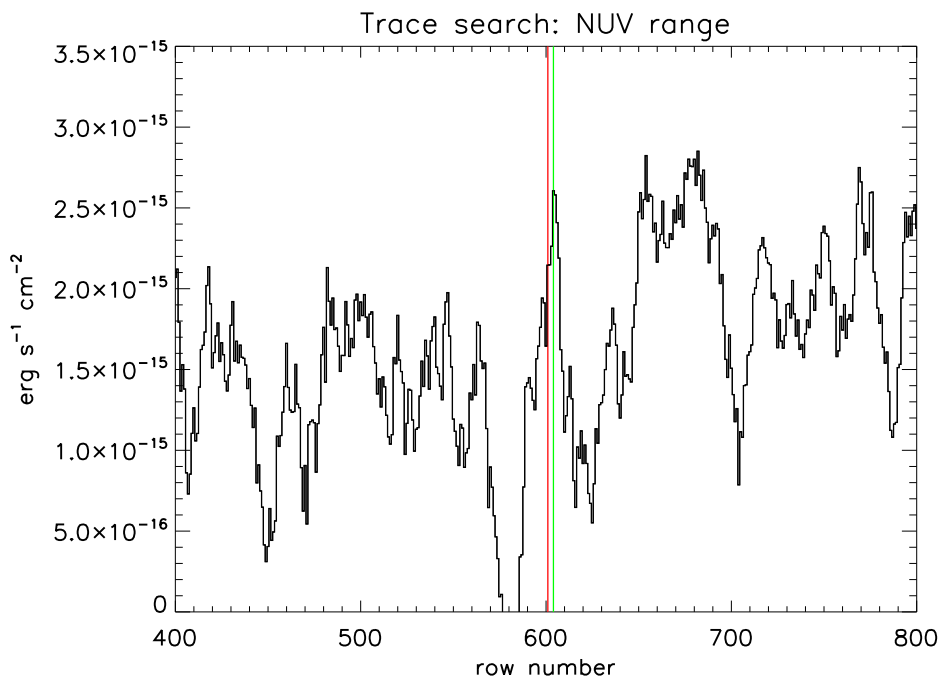}
   \caption{Trace search for NUV signal. Integrated fluxes along the
   axis of dispersion (along rows) as a function of row
   number. The black curves shows integrated fluxes within the
   wavelength range 2200 to 3110 \AA$\;$ and the blue curve fluxes
   within 3065 and 3110 \AA. The vertical red line indicates the row where
   the target is expected to be located and the green line indicates a shift
   of three rows where flux displays a local maximum. 
   }
              \label{f:NUV-Trace}%
    \end{figure*}
   \begin{figure*}
   \centering
   \includegraphics[width=12cm]{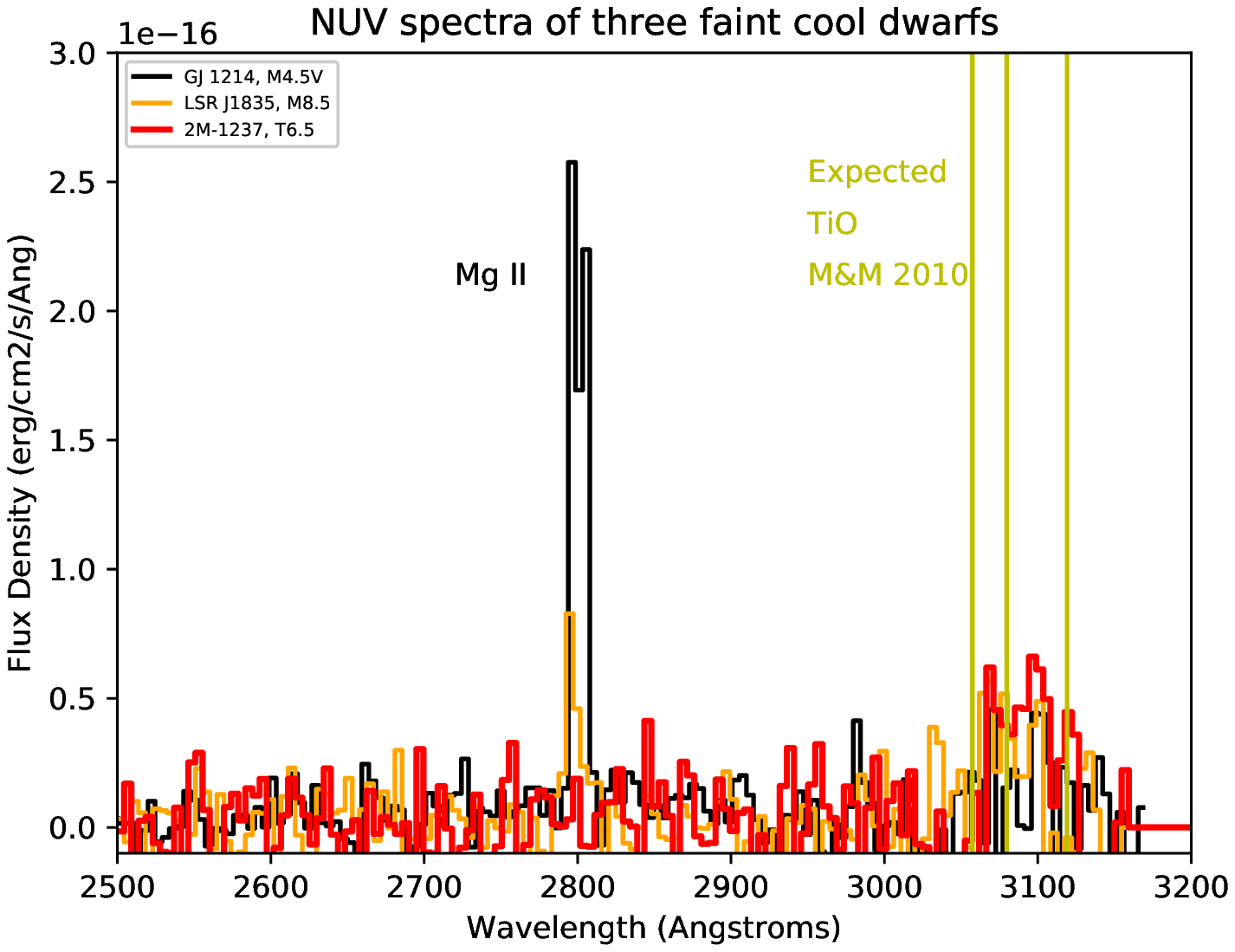}
   \caption{Near-UV spectrum of 2MASS J1237+6526 compared with the M8.5
   ultracool dwarf LSR J1835+3259 and M4.5 low mass star GJ 1214. Around
   2800 \AA, the Mg II doublet is visible. The vertical yellow lines show
   TiO bands based on \cite{mili10}.}
   \label{f:NUV_compare_ucd}%
    \end{figure*}
%
The NUV observations were executed with STIS grating
G230L for the wavelength range 1570–3180 \AA.  The resultant spectra
obtained from the two exposures and the combined spectrum are shown in
Figure \ref{f:NUV_Spectrum} as black lines. 
The light blue lines show the uncertainty spectra based on the
variance of the background fluxes (Appendix \ref{a:data_analysis}).
At the lower end and somewhat at the higher end of the wavelength range the spectrum is noisy,
similar to other HST/STIS observations with G230L resulting from the decreased
sensitivity (see STIS Instrument Handbook).

The signal in the NUV is generally very faint and
barely exceeds the
noise level.
Integrating between 2200 and 3000 \AA, we find a flux of
1.6 $\times$ 10$^{-15}$ erg s$^{-1}$ cm$^{-2}$  with a S/N of
2.4. Only within 3065 - 3110 \AA$\;$ is a  distinct feature visible (within the two red lines in Figure \ref{f:NUV_Spectrum}). It possesses a net
flux of 2.0 $\times$ 10$^{-15}$ erg s$^{-1}$ cm$^{-2}$  with a S/N of
4.8.
We note, however,  that the emission occurs near the long wavelength end of the
grating G230L where the spectrum appears very noisy.
Based on the counts of the signal and the background counts in the
flt-files, we find a S/N
of 3.8  within 3065-3110 \AA.
The counts over the lower wavelength
range within 2200 and 3000 \AA\;
are likely not significant because the net counts and resulting
S/N can have values higher or lower than one, depending on the
details of choosing the rows to obtain the net signal and the rows for
determining the background.

To further investigate the significance of the  NUV signal from the
target, we  search for a signal by integrating  along the axis of dispersion for each
row within
the unprocessed x2d-files (i.e., without subtracting background emission).
The results are shown in Figure \ref{f:NUV-Trace} for integration within the
wavelength range 3065 and 3110 \AA$\;$. The vertical red
line in Figure \ref{f:NUV-Trace} shows the row where the target is
expected to be located, and the green line is shifted by three rows to
better fit the local maximum of the flux. This small shift lies within
the pointing uncertainties.
A localized enhancement of
integrated flux within 3065-3110 \AA$\;$ is visible near the row where the target is
expected to be located.
Its amplitude is consistent with the derived
flux of 1.6 $\times$ 10$^{-15}$ erg s$^{-1}$ cm$^{-2}$.
However, Figure \ref{f:NUV-Trace} also shows that the background is
highly inhomogeneous with
other peaks of similar amplitudes.
This cautions against a claim of an unambiguous detection of a NUV signal
from the source.

The origin of the
possible
net emission 
within 3065 - 3110 \AA$\;$
is not
clear, but it could stem from emission of TiO possibly excited by
auroral electrons
\citep{palm72,phat70,mili10}. Low S/N emission in the same
wavelength range was also observed in the spectra of the ultracool M8.5 dwarf LSR
J1835+3259 \citep{saur18}. The same spectral region is displayed in
more detail in Figure \ref{f:NUV_compare_ucd}, where we also show
three expected TiO vibrational bands (yellow vertical lines) at  3057.4, 3079.7,
and 3119.1 \AA$\;$ calculated based on data in \cite{mili10}. While
the lower two wavelength bands only approximately coincide with peaks
in the observation, the band at  3119.1 \AA$\;$ fits closer to an
observed peak (red line). The wavelengths of the lower bands, however, are  also
less certain compared to the band at  3119.1 \AA$\;$ \citep{mili10}.
For example, the band at 3079.7 \AA$\;$ originates from the $'\nu$ = 2
level of G $^3$H
state which would require a favorable
Franck-Condon factor for excitation.
We note that
TiO emission from the upper atmosphere of a
T dwarf would be surprising since TiO is
expected to be depleted in the L-dwarf stage and rained out as
minerals \citep[e.g.,][]{burr00,reid13}.

All in all, the analysis shows that the low S/N feature between 3065 - 3110 \AA$\;$ within the inhomogeneous
background flux
needs to be taken with caution.
Keeping this uncertainty in mind, we compare the 
NUV spectrum of the T-type brown dwarf
2MASS J1237+6526 
with slightly more massive dwarfs,  such as
the ultracool M8.5 dwarf LSR J1835+3259 \citep{saur18} and the 
M4.5V star GJ 1214 \citep{fran16}. The spectra of these three targets
are displayed in Figure \ref{f:NUV_compare_ucd}.
Fluxes associated with
\ion{Mg}{II} at 2796 and 2803 \AA$\;$ 
\citep[e.g.,][]{feld96,fran13}
are not detectable within the
spectrum of 2MASS J1237+6526, but grow with increasing mass of the
targets. The trend of increasing \ion{Mg}{II} luminosity with mass holds
on average also for more massive M and K dwarfs observed within the MUSCLES
program by \citet{fran13,fran16}.
The tentative emission possibly stemming from TiO within
3065 and 3110 \AA$\;$ is
largest for 2MASS J1237+6526, which is  the coldest and lowest mass object of
the three. The flux in the same band is lower for LSR J1835+3259,
while it appears negligible for GJ 1214.

\subsection{Far-UV spectrum}
   \begin{figure*}
   \centering
   \includegraphics[width=15cm]{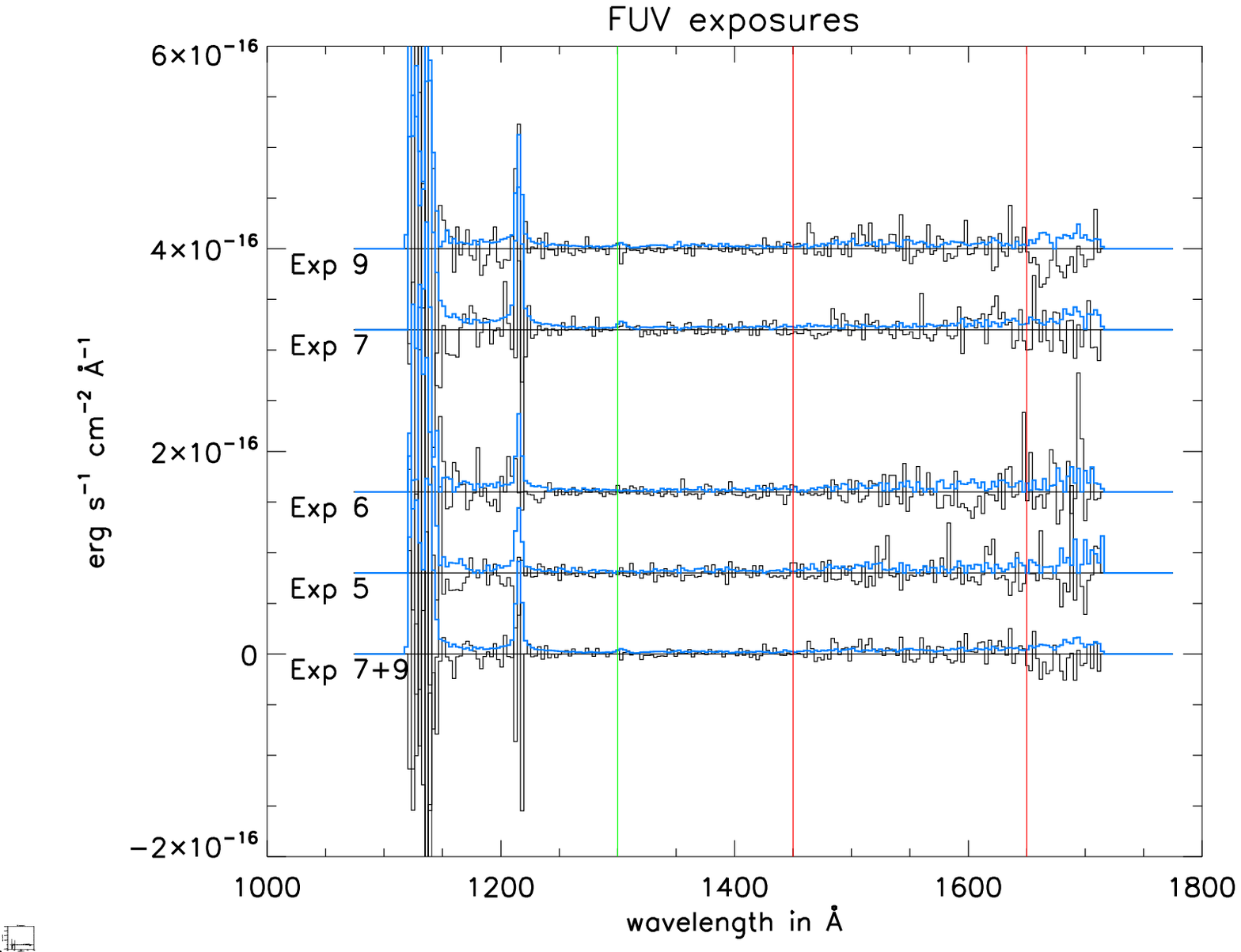}
   \caption{Far-UV spectra of 2MASS J1237+6526. The spectra are shifted by a
   constant offset for display purposes. The red lines at 1450
   and 1650 \AA$\;$ and the green line at 1300 \AA$\;$ indicate the region within
   which integrated fluxes are calculated (see text). In light blue we display
   the uncertainty spectrum calculated with the variance method described in
     Appendix \ref{a:data_analysis}.}
   \label{f:FUV_SPECTRUM}%
    \end{figure*}
%
   \begin{figure*}
   \centering
   \includegraphics[width=15cm]{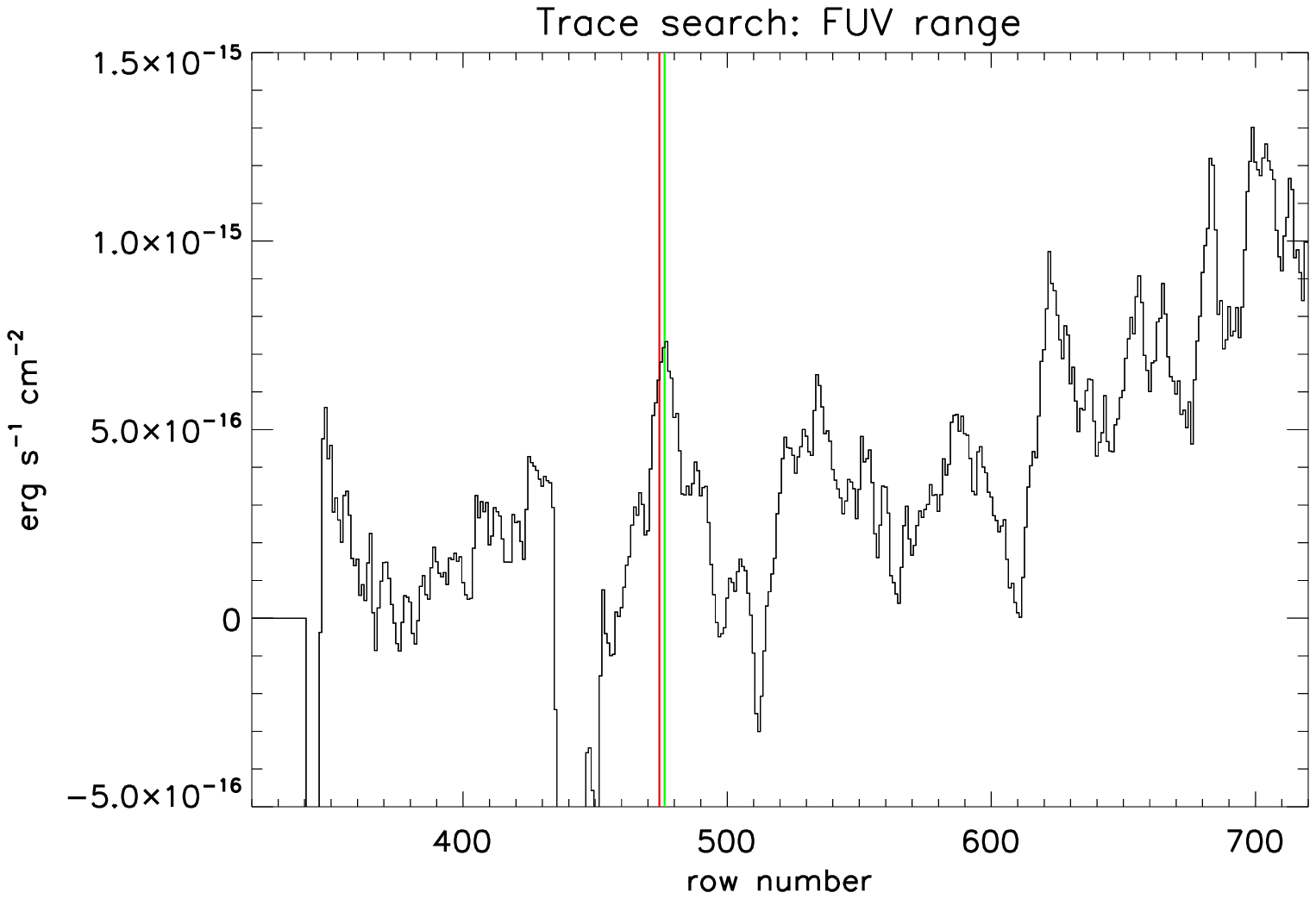}
   \caption{Trace search at FUV wavelength for 2MASS J1237+6526. Integrated
   fluxes along rows within 1450 and 1650 \AA\ (i.e., along the
   axis of dispersion), displayed as a function of row
   number  (exposures 7 and 9 combined).
   The vertical red line indicates the row where
   the target is expected to be located, and the green line indicates a shift
   of two rows where the flux displays a local maximum.}
   \label{f:FUV-Trace}%
    \end{figure*}
%
%
The FUV observations were taken with STIS grating G140L within the
wavelength range 1150-1730$\;$ \AA$\;$ (see Table \ref{table:1}). The
results of each exposure and exposures 7 and 9  combined are shown in
Figure \ref{f:FUV_SPECTRUM}.
The light blue lines show the residual spectra based on the
variance method (Appendix \ref{a:data_analysis}).

The brown dwarf is overall extremely faint in the FUV. The only
wavelength range
with somewhat enhanced fluxes lies within 1450 and
1650 \AA.
Between 1300
and 1450 \AA$\;$ the fluxes are negligible. For wavelengths 
shorter than 1300 \AA$\;$  and longer than
1650 \AA$\;$ the sensitivity is low compared to the mid-wavelength range and
the resultant spectrum turns noisy without
significant flux
consistent with the uncertainty spectrum in 
Figure \ref{f:FUV_SPECTRUM}.
The flux within 1450 and
1650 \AA$\;$
differs
 between the four exposures. It
is not significant for exposures 5 and 6 taken during
visit 2 with a S/N  < 1. Exposures 7 and
9 taken during visit 2 lead
to an integrated flux of 4.0 $\times$
10$^{-16}$ erg s$^{-1}$ cm$^{-2}$  with a  S/N of 2.6 and 
7.8 $\times$
10$^{-16}$ erg s$^{-1}$ cm$^{-2}$  with a S/N of 4.6, respectively.
A S/N estimate based on counts
in the flt-flies within
1450 and 1650 \AA$\;$ (see Appendix \ref{a:data_analysis})
for exposure  9 leads to a S/N of 3.5.

To independently assess the significance of the fluxes from the target
for exposures
7 and 9 combined, we integrate the fluxes within each row (i.e., along the
direction of dispersion) within 1450 and 1650 \AA$\;$ for each
row. These integrated fluxes are displayed as a function of row number
in Figure \ref{f:FUV-Trace}.
The vertical red line indicates the row where the target is expected to be located,
which coincides with a local maximum of the observed flux. The
green line is shifted by two rows, which slightly fits better and which
is used as target location for calculating the fluxes in this
subsection. This small shift lies within the pointing uncertainties.
Figure \ref{f:FUV-Trace}
shows both a wide variability and a trend in the background fluxes
preventing a unique detection.

Possible
emission between 1450 and 1650 \AA$\;$ would be consistent
with expectations from H$_2$ Lyman and Werner bands as observed, for
example,
with the Galileo
Ultraviolet Spectrometer and the Hopkins Ultraviolet Telescope
\citep{pryo98,wolv98}. In particular, peak fluxes from 2MASS
J1237+6526
at wavelength bands slightly longer  than
1600 \AA$\;$ are also observed and theoretically expected from Jupiter's
auroral emission \citep{wolv98}. Expected smaller amplitude features for lower FUV
wavelengths from H$_2$ Lyman and Werner bands might be too weak to be
resolved in the dwarf spectrum.
The possible H$_2$ emission cannot stem from the geocorona because of
the
extremely low H$_2$ number densities at and above HST altitudes due to
the rapid oxidation of H$_2$ and conversion to H atoms in the Earth's
upper atmosphere below 200 km. This is consistent with all previous
analyses of HST observations at these wavelengths.

\subsection{Ly-$\alpha$ spectrum}
%
   \begin{figure*}
   \centering
   \includegraphics[width=15cm]{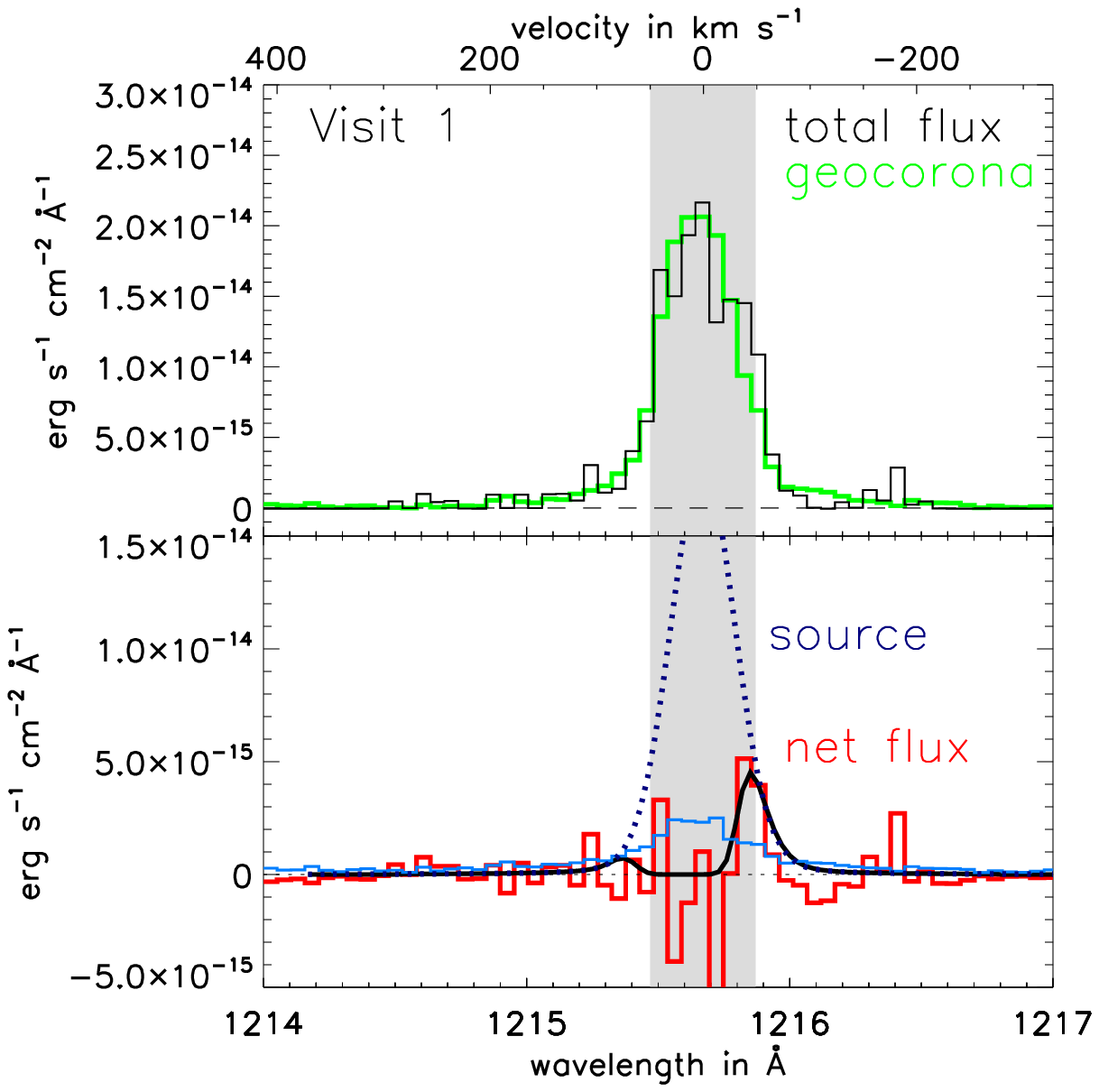}
   \caption{Ly-$\alpha$ spectrum of Visit 1 of 2MASS J1237+6526. The top
   panel shows total flux and geocoronal emission within the slit determined from rows
   away from the target. The bottom panel displays the net flux from the
   target and the possible source flux (see text for  details).
   In light blue we display
   the uncertainty spectrum calculated with the variance method described in
   Appendix \ref{a:data_analysis}.}
   \label{f:Lya_SPECTRUM_V1}%
    \end{figure*}

   \begin{figure*}
   \centering
   \includegraphics[width=15cm]{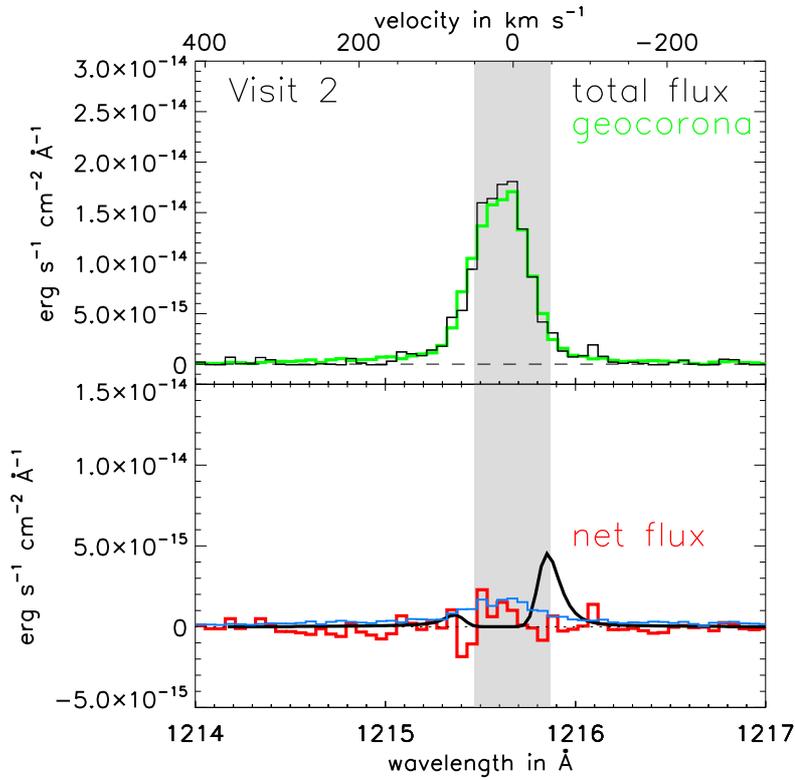}
   \caption{Ly-$\alpha$ spectrum of Visit 2 of 2MASS J1237+6526. Labels are the same as in  Figure \ref{f:Lya_SPECTRUM_V1}. }
   \label{f:Lya_SPECTRUM_V2}%
    \end{figure*}
   \begin{figure*}
   \centering
   \includegraphics[width=15cm]{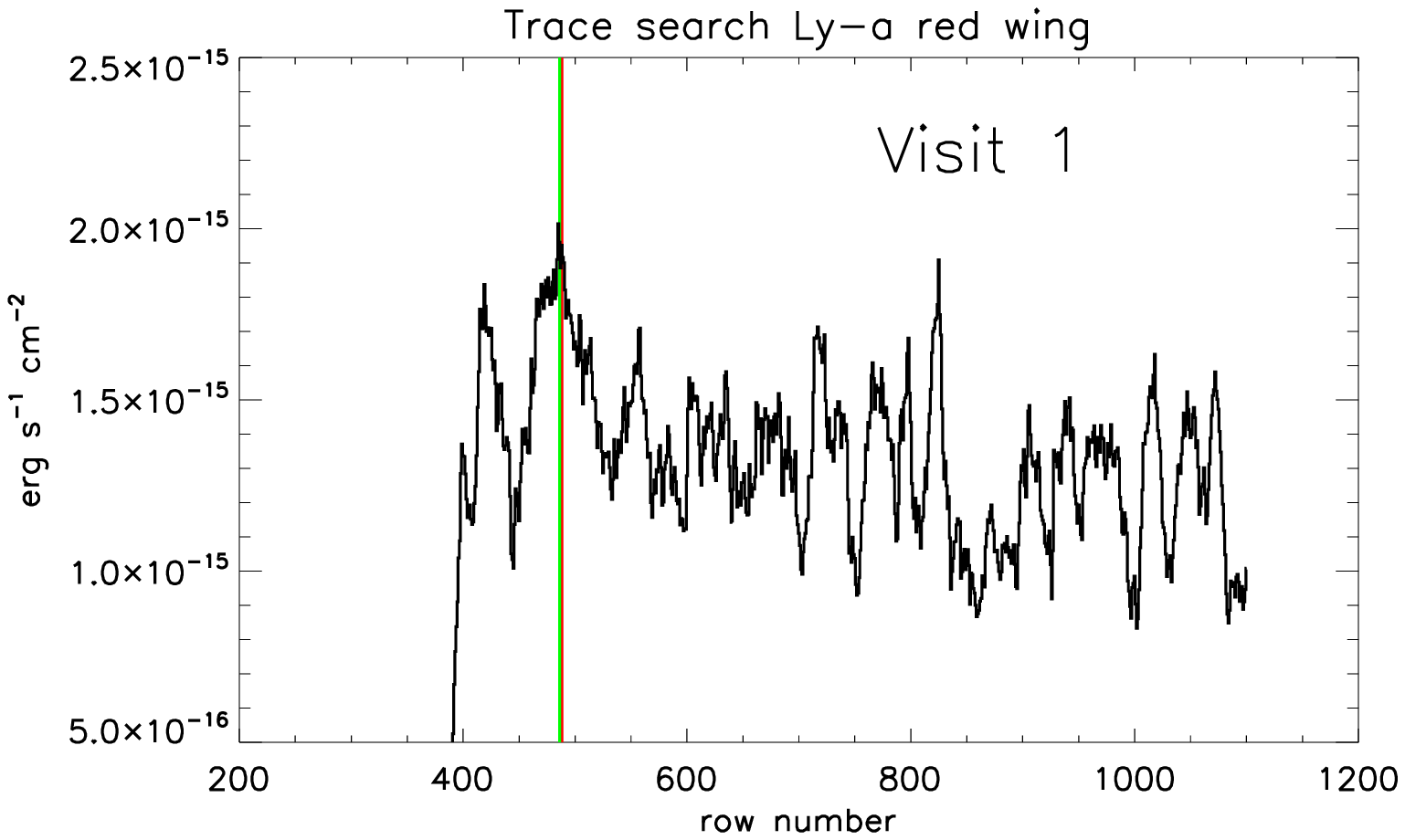}
   \caption{Trace search at Ly-$\alpha$ redwing for 2MASS
   J1237+6526. }
   \label{f:Lya-Trace}%
    \end{figure*}
%
%
Lyman-$\alpha$ emission is searched for with STIS grating
G140M at central wavelength 1218 \AA. In Figures
\ref{f:Lya_SPECTRUM_V1} and \ref{f:Lya_SPECTRUM_V2}, we display the
Ly-$\alpha$ spectrum for visit 1 and visit 2, respectively. The top
panels in both figures show the total Ly-$\alpha$ fluxes in black for
rows where the target is located. The averaged Ly-$\alpha$ flux in
rows away from the target is shown in green. These fluxes result from
the geocoronal emission, and need to be subtracted to retain the net
fluxes from the target. These net fluxes are displayed in the lower
panels of Figures \ref{f:Lya_SPECTRUM_V1} and \ref{f:Lya_SPECTRUM_V2}
   together with the uncertainty spectrum in light blue based on the
   variance method in the Appendix. We note, in the case of larger background
   fluxes such as geocoronal emission, that it underestimates the
   uncertainty, and that this effect leads to lower S/N values as
   those
based on counts (see discussion below). Therefore larger negative outliers of the net flux are 
   possible at wavelengths with large geocoronal emission (displayed
   as a green line).
Due to absorption in the interstellar medium, emission near the
central Ly-$\alpha$ wavelength 1215.67 \AA$\;$ is not detectable
\citep{lins96,lins14,fran13}. The spectral range
where the geocorona is dominant is shown as a gray
shaded area in both figures. The net emission observed for visit 1
shows a small surplus in the Ly-$\alpha$ red wing longward of 1215.8
\AA. The net flux within the wavelength range 1215.8 to 1216.5 \AA$\;$ is
4.9 $\times$ 10$^{-16}$ erg s$^{-1}$ cm$^{-2}$ with a S/N of 3.7
based on the variance method described in the Appendix.
In contrast, visit 2
displayed in Figure \ref{f:Lya_SPECTRUM_V2}, shows no sign of a surplus
in either the red or the blue wings. Due to large background counts
from the geocorona, the S/N calculations based on the variance methods
overestimate the S/N.
The large background flux is considered in the S/N calculations based
 on the counts in the flt-files. This therefore leads to a lower S/N
  of 1.9.
We note that we chose to subtract the background from the trace
flux within the x2d-files because small deviations in the not-fully-rectified flt-files lead to spatial
uncertainties in the background count structure, which can introduce
significant errors in the subtraction of these files.

The
possible 
surplus in the red wing of visit 1 is further investigated in
Figure \ref{f:Lya-Trace}, where we integrate the flux (without any
background subtraction) in each row between 1215.8 and 1216.5 \AA$\;$ and show this
flux as a function of row number. The row where the target is expected
to be located is shown as red vertical line in Figure
\ref{f:Lya-Trace}. It is colocated with the maximum flux demonstrating
independently a small possible net surplus within the Ly-$\alpha$ red wing. The local
maximum appears to be slightly shifted 2 rows downward (green
line in the figure). Such a small offset is typical within the pointing
uncertainties.
Figure \ref{f:Lya-Trace} shows that the background emission possesses
wide variability including
a few
other large peaks of nearly similar height,
which introduces uncertainty
in the unique detection of
a signal from the source.

The possible
net flux in Figure \ref{f:Lya_SPECTRUM_V1} can  be used to roughly
estimate the underlying Ly-$\alpha$ source emission. We apply the
same procedure as in \citet{saur18} based on reconstruction techniques
from  
\citet{wood05}, 
\citet{fran13}, 
\citet{bour15}, and 
\citet{youn16}, 
among others.
Since the observed
emission is very noisy we refrain from applying a quantitative
inversion to obtain the source function and unconstrained parameter
of the interstellar medium. Instead we manually explore parameter
space and present a reasonable fit with the aim of having an order of
magnitude constraint of the initial source luminosity.
Therefore, the source spectrum is
modeled with a Voigt profile with a Doppler width of 40 km s$^{-1}$,
a damping parameter of 0.09, and a peak flux of 2 $\times$ 10$^{-14 }$
erg s$^{-1}$ cm$^{-2}$ \AA$^{-1}$. The absorption in the interstellar
medium by hydrogen \ion{H}{I} and deuterium \ion{D}{I} is described
using a Lorentzian
absorption with cross-sections and ratios given in \citet{saur18}.
We assume a hydrogen column
density of the interstellar medium of 1 $\times$  10$^{17}$ cm$^{-2}$
consistent with estimates from the LISM calculator \citep{redf08}.  The
radial velocity of the brown dwarf 2MASS J1237+6526 is unknown and the
velocities of the interstellar medium are only a few km s$^{-1}$  based on
the LISM kinetic calculator \citep{redf08}. For a reasonable fit to
the data we assume a relative velocity between the source of the
emission and the ISM of 16 km  s$^{-1}$. However, velocities of 8 km
 s$^{-1}$ and a column density of  2 $\times$  10$^{16}$ cm$^{-2}$ or
 a velocity of  10 km  $s^{-1}$ or 1 $\times$  10$^{16}$ cm$^{-2}$
 give equally good fits. 
The resultant Ly-$\alpha$
profile is convolved with the line spread function from the STIS
Instrument Handbook \citep{rile17}. In
Figure \ref{f:Lya_SPECTRUM_V1} we show the resultant fit by eye as black
curve. The underlying Ly-$\alpha$ source spectrum from the brown dwarf 2MASS
J1237+6526 is displayed as a dashed line in the figure. Integrating the
source Ly-$\alpha$ flux derived for visit 1 leads to a total flux of
5.8 $\times$ 10$^{-15 }$ erg s$^{-1}$ cm$^{-2}$. 

Only for guiding the eye, we display in Figure
\ref{f:Lya_SPECTRUM_V2} the fit to the net flux from visit 1  (black
line).
This demonstrates that no net Ly-$\alpha$ was detected during visit 2 in
excess of the noise level.

\subsection{Band luminosities of target}
%
   \begin{figure*}
   \centering
   \includegraphics[width=15cm]{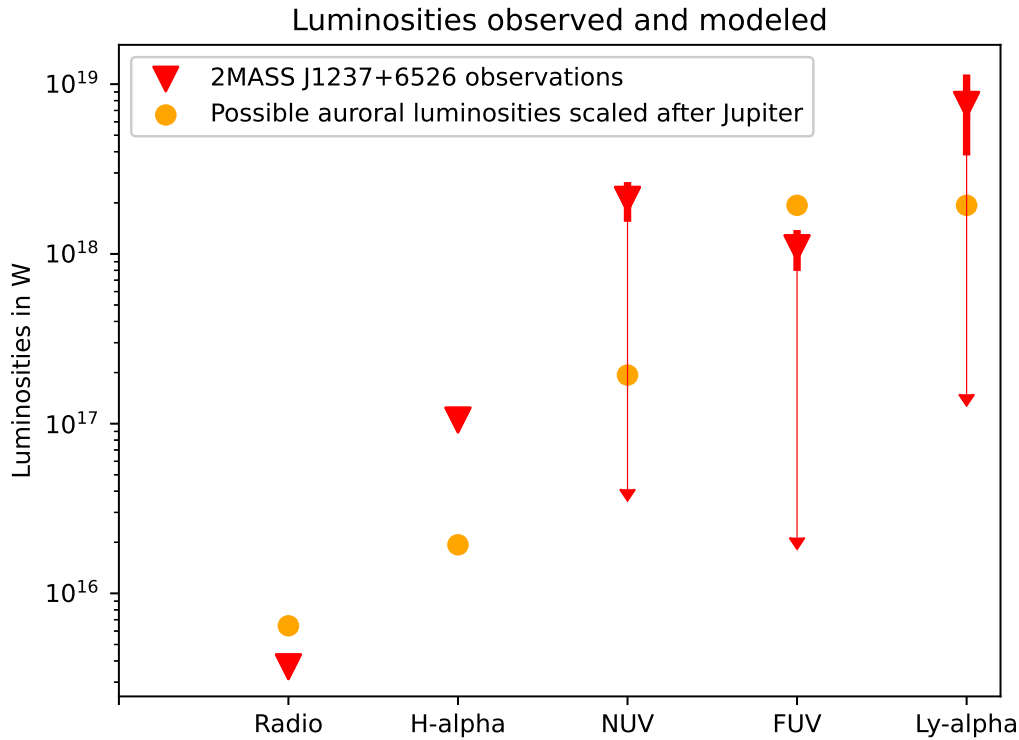}
   \caption{Possible luminosities of brown dwarf 2MASS J1237+6526 (red
   triangles) compared with
   theoretically derived auroral luminosities (orange
   circles). Auroral luminosities are scaled after Jupiter
   and are calculated with
   L$_{Jupiter}$ S$_{pot,rel}$ Q$_{rel}(=1)$ (see text for details). The
derived luminosities need to be considered with caution,
but can be considered upper limits in any case, which is indicated
symbolically as
 downward 
arrows.
}
   \label{f:band_luminosities}%
    \end{figure*}
%
%
\begin{table*}
      \caption[]{Possible values and upper limits for the UV luminosities
      of 2MASS J1237+6526 (see text).}
\label{t:band_luminosities}      
\centering          
\begin{tabular}{l c r r r r l  l l}     
\hline\hline       
type &wavelength band & flux & S/N$^a$ & Luminosity\\ 
\hline\
 & \AA  &erg s$^{-1}$ cm$^{-2}$ & &watt & \\
\hline
NUV exposures \#3 + \#4 & 3065 - 3110 &1.6 $\times$ 10$^{-15}$ & 3.8 & 2.1 $\times$
10$^{18}$ \\
FUV exposure \#9 &  1450 - 1650  &7.9 $\times$ 10$^{-16}$ & 3.5 & 1.0 $\times$ 10$^{18}$
\\
Ly-$\alpha$ exposures \#1 + \#2 &  1215.8 - 1215.5  & 6.2
$\times$ 10$^{-16}$ & 1.9 & 7.6 $\times$ 10$^{18}$ \\
\hline                    
\hline                  
\end{tabular}
\tablefoot{$^a$: Lowest value of S/N from the methods based on
variances or counts (see main text).}
\end{table*}

With the possible
fluxes derived in the previous subsections, we calculate
possible
luminosities for the associated wavelength bands. They are summarized
in Table \ref{t:band_luminosities} and displayed in Figure
\ref{f:band_luminosities} as red triangles. The formal S/N in the
observations are shown with the red vertical bars around the
triangle.
As the
derived values are subject to additional
uncertainty due to the background subtraction which introduces
systematic uncertainties that are difficult to fully assess, the
derived luminosities need to be considered with appropriate caution,
but can be considered upper limits in any case. We indicate this
symbolically with
the thin downward 
arrows.
 Luminosities at H-$\alpha$ and radio
wavelengths are taken from the literature \citep{burg02,kao18}.
For comparison we show expected auroral emission with the scaling model
derived in Section \ref{s:theory}.
The orange circles show the expected
auroral luminosities based on expression (\ref{e:P_mag}) with the auroral
luminosities from Jupiter L$_{mag,J}$, the auroral power potential $S_{pot,rel}$ of 2MASS
J1237+6526 relative to Jupiter and assuming $Q_{rel} = 1$ (i.e., that the
auroral generator within the magnetosphere of 2MASS
J1237+6526 is similar to that of  Jupiter).

\section{Discussion and conclusion}
\label{s:discussion}

In this section we discuss further
under what circumstances the possible UV emission from 2MASS J1237+6526 could originate from auroral
processes by using the expressions for auroral powers from Section
\ref{s:scaling} for a mass flow driven or companion driven aurora, respectively.
Our analysis of HST observations shows that the derived UV
luminosities of 2MASS J1237+6526,
if confirmed, lie in the range 
10$^{18}$ to 10$^{19}$
watt in the NUV, FUV, and at Ly-$\alpha$ wavelengths. They otherwise 
provide upper limits.
Jupiter's total auroral UV emission $P_{Jupiter}$ lies in the
range 2--10 $\times$
10$^{12}$ watt and is thus a factor of $\sim$10$^6$ smaller.
Based on the derivations in Section \ref{s:theory},
the auroral
power potential
of Jupiter is S$_{pot,J}$ = 600 watt m$^{-2}$
Siemens$^{-1}$ and the power potential of 
2MASS J1237+6526 is $S_{pot,2M}$ = 3.9 $\times$ 10$^8$ watt m$^{-2}$
Siemens$^{-1}$. This leads to a relative power potential of
$S_{pot,rel} =  S_{pot,2M} / S_{pot,J}$ = 6.5 $\times$ 10$^5$. Thus, the
relative power potential differences
could
explain the large difference
between Jupiter's UV emission and the possible UV emission from 2MASS
J1237+6526. This is also shown in Figure \ref{f:band_luminosities},
where we  use the average values of the UV luminosities 
for Jupiter from Table \ref{table:epsilon} combined with the scaling
model from expression (\ref{e:P_mag}). 
The comparison shows that within an order of magnitude the auroral
luminosities and possible UV emission from 2MASS J1237+6529 could be
comparable.

Alternatively, a planetary companion can explain the
possible UV
emission as well. This scenario is described by expression (\ref{e:P_companion}).
The total UV luminosity of Io's northern and
southern auroral spots $P_{Io}$ combined lies in
the range of 1-10 $\times$ 10$^{10}$ watt 
\citep{pran96,wann10,bonf13},  about a factor of 100 smaller
compared to Jupiter's main auroral emission driven by mass
transport.  The same power potential
is applicable for both types of aurora, which means that the luminosity in (\ref{e:P_companion}) can render values
comparable to the observation with $Q_{rel}$ = 100. This is possible, for
example,  for a planetary companion at the same radial
distance as Io from Jupiter (5.9 R$_J$), but with a radius 10
times larger than Io's ($R_{Io}$ = 1800 km),
corresponding to 2.8 times the radius of Earth,
or, alternatively, an Earth-sized planet at a distance of 3.5 R$_J$. Earth-sized planets around
brown dwarfs are thought to be possible \citep{payn07}, but no short-period companion around 2MASS J1237+6526 has been reported as transits
in photometric surveys to the authors' knowledge.

Both auroral scenarios would be consistent with
the observed H-$\alpha$ emission from the dwarf
\citep{burg02,lieb07}. They could also be the
driver of the radio emission reported in \citet{kao16,kao18}.

Due to the higher mass of 2MASS
J1237+6526 compared to that of  Jupiter by a factor of about 30, the orbital period
of a companion located at a distance similar to Jupiter's
moon Io would be shorter by a factor of 5.5 with a rotation
period of 
7.8 hr.
Depending on the location where the coupling
of a planetary companion intersects the surface of the brown dwarf,
Doppler shifts of the emission due to the rotation of the dwarf 
on the order of 10 km s$^{-1}$ could emerge.
This would contribute to the Doppler shift of the Ly-$\alpha$
emission. It might also allow 
to identify the source
by observations of periodic oscillations of the shift in future measurements.

Any variability of the possible auroral UV emission from the brown dwarf during
the different FUV and Ly-$\alpha$ visits could
in principle be explained  by a planetary companion whose
auroral effects are obstructed (or partially obstructed)
because they lie behind the dwarf as
seen from Earth (see discussion in e.g.   \citealt{fisc19}). However, 
tilted magnetic fields of the central body can cause time variability
as well. Due to the
uncertainty of the significance of
the HST observations, it is
premature
to explore the various possibilities further. We note that
H-$\alpha$ emission was observed to be persistent within a factor of
two \citep{burg02}.

An important question is whether the H-$\alpha$ emission and the
possible UV
emission from 2MASS J1237+6526 could be caused by chromospheric
emission. \citet{lieb07} argue that this brown dwarf does not possess a very active
chromosphere. One of their arguments is based on the very faint
absolute
J-band
magnitude of 15.88
$\pm$ 0.13 \citep{vrba04},
which is similar to other T6-7 dwarfs and rules
out that 2MASS J1237+6526 is overluminous. In addition, the kinematics
of the objects measured by \citet{vrba04} suggest that it is not a
young object. Finally, its near-IR spectrum provides evidence that
2MASS J1237+6526 has a high surface gravity and/or a low metallicity,
which implies that the dwarf is old
\citep{lieb07}.
These arguments speak against an
overluminous chromosphere as the source of the H-$\alpha$ and UV
emission.
However, chromospheric (i.e., internally driven) activity is 
not very well understood for L- and T-type brown
dwarfs. Therefore, it remains an open question whether observed
emission at H-$\alpha$, radio, or possibly UV wavelengths
from 2MASS J1237+6526 is externally driven by auroral
acceleration process or by internal heating due to Alfv\'en waves, for example
\citep[e.g.,][]{rodr18}.

In summary, the possible UV emission of the brown dwarf 2MASS J1237+6526 are
consistent with theoretically expected auroral emission 
driven in the dwarf's magnetosphere by mass transport or a planet-sized companion. However, the observational results of this analysis
are
tentative and need to be considered with caution
due to the low S/N in combination with the nonsystematic background
variations.
We also did not detect uniquely identifiable spectral
lines and bands in the NUV and FUV spectrum  (except possibly
Ly-$\alpha$).
Further observations are required to establish the significance
of the UV emission.
Resolving time variability could constrain the auroral generator, and
thus allow 
to determine whether the emission is due to an external
source, due to auroral particle acceleration, in contrast to
internally produced plasma heating including internally driven
electron beams.
Despite the
tentativeness of the presented observations, brown dwarfs are ideal
objects to search for UV aurora outside of the Solar System, and
further UV observations might
prove greatly helpful in understanding
the space plasma environment around 
%
brown dwarfs and extrasolar planets.

\begin{acknowledgements}
This work is based on observations with the NASA/ESA
Hubble Space Telescope obtained at the Space Telescope
Science Institute, which is operated by the Association of
Universities for Research in Astronomy (AURA), Inc., under
NASA contract NAS 5-26555. We highly appreciate the help of Michael
Leveille in scheduling the observations and the valuable comments of
Johns Debes and his team on the data analysis.
This project has received funding from the European Research Council
(ERC) under the European Union’s Horizon 2020 research and innovation
programme (grant agreement No. 884711). DFS was supported by NASA through Grant HST-GO-15870.002-A
from the Space Telescope Science Institute. This research has made use
of the SIMBAD database, operated at CDS, Strasbourg, France. This work
has made use of data from the European Space Agency (ESA) mission {\it
Gaia} (\url{https://www.cosmos.esa.int/gaia}), processed by the {\it
Gaia} Data Processing and Analysis Consortium (DPAC,\url{https://www.cosmos.esa.int/web/gaia/dpac/consortium}). Funding for the DPAC has been provided by national institutions, in particular the institutions participating in the {\it Gaia} Multilateral Agreement.
\end{acknowledgements}

\begin{appendix}
\section{References and details of objects for auroral power potential
calculations}
\label{a:2}
Table \ref{t:objects} lists the objects for which magnetic fields,
rotation rates, and auroral power potential are  discussed in
Section \ref{s:scaling}. The references used in Table \ref{t:objects} are as follows: \\
(1): \citet{show99},
(2): \citet{seid02},
(3): \citet{deli98}, 
(4): \citet{kive02}
(5): NASA/JPL, retrieved from $\mbox{https://ssd.jpl.nasa.gov/?planet\_phys\_par}$,
(6): NASA/JPL, retrieved from $\mbox{https://solarsystem.nasa.gov/resources/681/solar-system-temperatures/}$,
(7): \citet{ande11},
(8):  \citet{theb15},
(9): \citet{conn93},
(10): \citet{doug18},
(11): \cite{conn18},
(12): \cite{turn21},
(13): \cite{brog12},
(14): \cite{kao16},
(15):  \cite{kao18},
(16): \citet{berd17},
(17): \citet{hall08},
(18): \citet{mori10},
(19): \citet{lins95}, 
(20): \citet{rein07},
(21): \citet{vang18},
(22): \citet{luge17},
(23): \citet{delr18}, 
(24): \citet{casa08}, 
(25): \citet{gaia18}, 
(26): \citet{pavl06}, 
(27): \citet{rein10},
(28): \citet{segr03},
(29): \citet{rein08a},
(30): \citet{segr03}, 
(31): \citet{lepi13}, 
(32): \citet{rein09},
(33): \citet{mori08a},
(34): \citet{mori08b},
(35): \citet{take07},
(36): \citet{saik16},
(37): \citet{bouc05},
(38): \citet{henr08},
(39): \citet{mout07},
(40): \citet{kerv16}.
(41): \citet{barn16},
(42): \citet{phan09},
(43): \citet{gaid14a}, 
(44): \citet{rabu19},
(45): \citet{rein18},
(46): \citet{gaid14b}, 
(47): \citet{roja12}, 
(48): \citet{vanb09}, 
(49): \citet{kova04},
(50): \citet{dona99},
(51): \citet{luck18}, 
(52): \citet{luck17}, 
(53): \citet{boya15}, 
(54): \citet{bain12}
(55):  \cite{froh07},
(56): \citet{clem04}, 
(57): \cite{basr88},
(58): \cite{peti08},
(59): \citet{gaul10}
(60): \citet{grie18}, 
(61): IAU 2015 resolution B3,
(62): \citet{jone04}, 
(63): \citet{bana98},
(64): \citet{mars06a},
(65): \citet{mars06b},
(66): \citet{foss13},
(67): \citet{bonf16},
(68): \citet{wrig04},
(69): \citet{kim96},
(70): \citet{blac94},
(71): \citet{bors15},
(72):  \citet{dona08},
(73): \citet{nort98},
(74): \citet{wade16},
(75): \citet{zore12},
(76): \citet{hubr09},
(77): \citet{nein03b},
(78): \cite{snow94},
(79): \citet{pasi01},
(80): \citet{hohl10}, 
(81): \citet{niev14},


The B-field values from planets, brown dwarfs, and stars
are obtained with different measurement
techniques. Details on the remote sensing methods can be found  in \citet{rein12}, among others. The magnetic field measurements from Table \ref{t:objects}
are based on the following methods:
1: In situ,
2: Radio ECM,
3: Bf (average magnetic flux from Stokes I),
4: <Bz> (average measured longitudinal field strength from Stokes V),
5: $Bz_{max}$ (maximum measured longitudinal field strength from Stokes V),
6: <B> (average field strength from ZDI map with Stokes V and I),
7: $B_{max}$ (maximum field strength from ZDI map with Stokes V and I),
d: $B_{max}$ (maximum field strength from ZDI map with Stokes V only),
e: $Bz_{max}$ (maximum measured longitudinal field strength from Stokes V, Q, and U).

\begin{table*}
\label{t:objects}
{\small
\begin{tabular}{c c r r r r r l}
\hline
Object name & Classification & M [kg] & R [m] & P [h] & $T_{eff}$ [K] & B [G] & B-meth$^f$\\
\hline
Ganymede & Moon & $1.43 \times 10^{23}$ (1) & $2.63 \times 10^{6}$ (2) & $171.71$ (2) & 110 (3) & $0.0144$ (4) & 1\\
Mercury & Planet & $3.30 \times 10^{23}$ (5) & $2.44 \times 10^{6}$ (2) & $1407.51$ (2) & 703 (6) & $0.004$ (7) & 1\\
Earth & Planet & $5.972 \times 10^{24}$ (5) & $6.37 \times 10^{6}$ (2) & $23.93$ (2) & 289 (6) & $0.6$ (8) & 1\\
Uranus & Planet & $8.68 \times 10^{25}$ (5) & $2.54 \times 10^{7}$ (2) & $17.24$ (2) & 78 (6) & $1.0$ (9) & 1\\
Neptune & Planet & $1.02 \times 10^{26}$ (5) & $2.46 \times 10^{7}$ (2) & $16.11$ (2) & 72 (6) & $0.9$ (9) & 1\\
Saturn & Planet & $5.683 \times 10^{26}$ (5) & $5.82 \times 10^{7}$ (2) & $10.66$ (2) & 135 (6) & $0.42$ (10) & 1\\
Jupiter & Planet & $1.898 \times 10^{27}$ (5) & $6.991 \times 10^{7}$ (2) & $9.92$ (2) & 165 (6) & $20$ (11) & 1 \\
$\tau$ Boo b & Exoplanet & $1.129 \times 10^{28}$ (13) & $8.0 \times 10^{7}$ (12) & $79.49$ (13) & 1650 (13) &$10.7$ (12) & 2 \\
2MASS J10430758+2225236 & BD L8 & $2.187 \times 10^{28}$ (14) & $6.292 \times 10^{7}$ (a) & $2.21$ (15) & 1012 (14) & $3900$ (15) & 2 \\
SDSS J04234858-0414035 & BD L7 + T2.5 & $2.983 \times 10^{28}$ (14) & $6.292 \times 10^{7}$ (a) & $1.47$ (15) & 1084 (14) & $3900$ (15) & 2 \\
SIMP  J01365662+0933473 & BD T2.5 & $4.375 \times 10^{28}$ (14) & $6.292 \times 10^{7}$ (a) & $2.88$ (15) & 1104 (14) & $3200$ (15) & 2 \\
2MASS J10475385+2124234 & BD T6.5 & $> 5.170 \times 10^{28}$ (14) & $6.292 \times 10^{7}$ (a) & $1.78$ (15) & 888 (14) & $5600$ (15) & 2 \\
2MASS J12373919+6526148 & BD T6.5 & $> 5.568 \times 10^{28}$ (14) & $6.292 \times 10^{7}$ (a) & $2.28$ (15) & 851 (14) & $4100$ (15) & 2 \\
2MASS J18353790+3259545 & $\ast$ M8.5 & $1.044 \times 10^{29}$ (16) & $1.468 \times 10^{8}$ (16) & $2.84$ (17) & 2800 (16) & $5100$ (16) & 2 \\
VB 10 & $\ast$ M8 & $1.591 \times 10^{29}$ (18) & $6.26 \times 10^{7}$ (18) & $19.2$ (68$^c$) & 2600 (19) & $1300$ (20) & 3 \\
Trappist-1 & $\ast$ M8 & $1.770 \times 10^{29}$ (21) & $8.42 \times 10^{7}$ (21) & $79.2$ (22) & 2511  (23) & $600$ (27) & 3 \\
GJ 3622 & $\ast$ M6.5 & $1.790 \times 10^{29}$ (18) & $7.65 \times 10^{7}$ (18) & $36$ (18) & 2450  (24) & $110$ (18) & 7 \\
VB 8 & $\ast$ M7 & $1.790 \times 10^{29}$ (18) & $6.96 \times 10^{7}$ (18) & $24$ (16$^c$) & 3299 (25) & $2300$ (20) & 3 \\
WX Uma & $\ast$ M6 & $1.988 \times 10^{29}$ (18) & $8.35 \times 10^{7}$ (18) & $18.72$ (18) & 2700 (24) & $4880$ (18) & 7 \\
DX Cnc & $\ast$ M6 & $1.988 \times 10^{29}$ (18) & $7.65 \times 10^{7}$ (18) & $11.04$ (18) & 2840 (20) & $220$ (18) & 7 \\
CN Leo & $\ast$ M5.5 & $1.988 \times 10^{29}$ (18) & $8.35 \times 10^{7}$ (18) & $48$ (16$^c$) & 2800 (26) & $2400$ (20) & 3 \\
GJ 1245 B & $\ast$ M5.5 & $2.386 \times 10^{29}$ (18) & $9.74 \times 10^{7}$ (18) & $17.04$ (18) & 3294 (25) & $580$ (18)  & 7 \\
Proxima Centauri & $\ast$ M5.5 & $2.446 \times 10^{29}$ (28) & $1.01 \times 10^{8}$ (28) & $1992$ (29) & 3042 (30) & $600$ (29) & 3 \\
GJ 1156 & $\ast$ M5 & $2.784 \times 10^{29}$ (18) & $1.11 \times 10^{8}$ (18) & $11.78$ (18) & 3110 (31) & $360$ (18) & 7 \\
GJ 1224 & $\ast$ M4.5 & $2.983 \times 10^{29}$ (18) & $1.18 \times 10^{8}$ (18) & $103.2$ (16$^c$) & 3300 (10) & $2700$ (20) & 3 \\
GJ 1154 A & $\ast$ M5 & $3.580 \times 10^{29}$ (18) & $1.39 \times 10^{8}$ (18) & $40.8$ (16$^c$) & 3040 (31) & $2000$ (32) & 3 \\
GJ 51 & $\ast$ M5 & $4.375 \times 10^{29}$ (18) & $1.53 \times 10^{8}$ (18) & $24.48$ (18) & 3346 (25) & $5020$ (18) & 7 \\
V374 Peg & $\ast$ M4 & $5.568 \times 10^{29}$ (33) & $1.948 \times 10^{8}$ (33) & $10.7$ (33) & 3432 (25) & $1300$ (34) & 7 \\
GJ 65 B & $\ast$ M6 & $2.376 \times 10^{29}$ (40) & $1.11 \times 10^{8}$ (40) & $5.52$ (41) & 3296 (25) & $5800$ (41) & 3 \\
GJ 65 A & $\ast$ M5.5 & $2.436 \times 10^{29}$ (40) & $1.15 \times 10^{8}$ (40) & $5.76$ (41) & 3337 (25) & $4500$ (41) & 3 \\
Gl 490 B & $\ast$ M4 V & $2.983 \times 10^{29}$ (42) & $3.27 \times 10^{8}$ (42) & $12.1$ (42) & 3266 (43) & $1800$ (42) & 7 \\
Gl 729 & $\ast$ M3.5 & $3.480 \times 10^{29}$ (44) & $1.426 \times 10^{8}$ (44) & $68.88$ (45) & 3162 (44) & $2200$ (20) & 3 \\
EQ Peg B & $\ast$ M4.5 & $4.971 \times 10^{29}$ (33) & $1.739 \times 10^{8}$ (33) & $9.70$ (33) & 3309 (25) & $1200$ (33) & 7 \\
YZ CMi & $\ast$ M4.5 & $6.164 \times 10^{29}$ (33) & $2.017 \times 10^{8}$ (33) & $66.62$ (33) & 3125 (46) & $3000$ (33) & 7 \\
EV Lac & $\ast$ M3.5 & $6.363 \times 10^{29}$ (33) & $2.087 \times 10^{8}$ (33) & $104.92$ (33) & 3400 (47) & $1500$ (33) & 7 \\
EQ Peg A & $\ast$ M3.5 & $7.755 \times 10^{29}$ (33) & $2.435 \times 10^{8}$ (33) & $25.46$ (33) & 3585 (25) & $800$ (33) & 7 \\
AD Leo & $\ast$ M3 & $8.352 \times 10^{29}$ (33) & $2.643 \times 10^{8}$ (33) & $53.76$ (33) & 3390 (47) & $1300$ (33) & 7 \\
61 Cyg A & $\ast$ K5 V & $1.321 \times 10^{30}$ (35) & $4.31 \times 10^{8}$ (35) & $856.8$ (36) & 4526 (48) & $20$ (36) & 7 \\
LQ Hya & $\ast$ K0 V & $1.591 \times 10^{30}$ (49) &  $5.57 \times 10^{8}$ (50) & $38.4$ (50) & 4812 (51) & $800$ (50) & 7 \\
HD 189733 & $\ast$ K2 & $1.631 \times 10^{30}$ (37) & $5.29 \times 10^{8}$ (37) & $286.87$ (38) & 4875 (53) & $40$ (39) & 7 \\
$\epsilon$ Eri & $\ast$ K2 V & $1.631 \times 10^{30}$ (54) & $5.009 \times 10^{8}$ (55) & $268.80$ (55) & 5156 (56) & $350$ (57) & 3 \\
HR 1099 & $\ast$ K1 IV & $1.790 \times 10^{30}$ (50) & $2.3 \times 10^{9}$ (50) & $67.2$ (50) & 3266 (52) & $800$ (50) & 7 \\
HD 190771 & $\ast$ G2 V C & $1.909 \times 10^{30}$ (58) & $6.957 \times 10^{8}$ (58) & $211.2$ (58) & 5834 (58) & $51$ (58) & 6\\
HD 46375 & $\ast$ K0 V & $1.929 \times 10^{30}$ (59) & $6.33 \times 10^{8}$ (59) & $1008$ (59) & 5355 (60) & $5$ (59) & $7^d$ \\
18 Sco & $\ast$ G2 Va B & $1.949 \times 10^{30}$ (58) & $6.957 \times 10^{8}$ (58) & $544.8$ (58) & 5791 (58) & $3.6$ (58) &  6\\
Sun & $\ast$ G2V & $1.988 \times 10^{30}$ (61) & $6.957 \times 10^{8}$ (1) & $609.12$ (2) & 5780 (62) & $10$ (63) & 1 \\
V$\ast$ V401 Hya & $\ast$ G8/K0(IV) E & $2.008 \times 10^{30}$ (58) & $6.957 \times 10^{8}$ (58) & $295.2$ (58) & 5802 (58) & $42$ (58) & 6\\
HD 171488 & $\ast$ G2 V & $2.108 \times 10^{30}$ (64) & $8.00 \times 10^{30}$ (64) &  $31.2$ (65) & 5800 (64) & $500$ (65) & 7 \\
HD 117207 & $\ast$ G7 IV-V C & $2.148 \times 10^{30}$ (66) & $7.861 \times 10^{8}$ (67) & $864$ (68) & 5681 (67) & $1.36$ (66) & 4 \\
HD 1817 & $\ast$  F8 V & $2.287 \times 10^{30}$ (69) & $8.21 \times 10^{8}$ (70) & $24$ (65) & 6126 (51) & $250$ (65) & 7 \\
HD 76151 & $\ast$ G2 V B & $2.466 \times 10^{30}$ (58) & $6.957 \times 10^{8}$ (58) & $492$ (58) & 5790 (58) & $5.6$ (58) & 6 \\
$\tau$ Boo A & $\ast$ F7 IV & $2.76 \times 10^{30}$ (71) & $9.878 \times 10^{8}$ (71) & $79.44$ (72) & 6399 (71) & $10$ (72) & 7 \\
a Cen & $\ast$ B7 IIIpv & $7.357 \times 10^{30}$ (73) & $3.958 \times 10^{9}$ (73) & $211.68$ (74) & 17700 (73) & $470$ (74) & 5\\       
$\epsilon$ Dor & $\ast$ B6 V & $8.57 \times 10^{30}$ (75) & $2.643 \times 10^{9}$ (76) & $> 271.00$ (75) & 13700 (76) & $64$ (76) & 4 \\
$\zeta$ Cas & $\ast$ B2 IV & $1.65 \times 10^{31}$ (77) & $4.104 \times 10^{9}$ (77) & $128.88$ (74) & 20426 (77) & $30$ (74) & 5 \\
V1046 Ori & $\ast$ B1.5 V & $1.988 \times 10^{31}$ (41$^b$) & $3.130 \times 10^{9}$ (79) & $21.6$ (74) & 23700 (80) & $2035$ (74) & 5 \\
V901 Ori & $\ast$ B2 V & $1.988 \times 10^{31}$ (78) & $4.591 \times 10^{9}$ (78) & $36.96$ (74) & 22000 (78) & $1310$ (74) & $5^e$\\
HD 64740 & $\ast$ B1.5 Vp & $1.988 \times 10^{31}$ (41$^b$) & $5.843 \times 10^{9}$ (79) & $31.92$ (74) & 23700 (80) & $660$ (74) & 5 \\
$\beta$ Cep &$\ast$ B1 IV & $2.426 \times 10^{31}$ (81) & $3.896 \times 10^{9}$ (81) & $288$ (74) & 27000 (81) & $110$ (74) & 5 \\
NU Ori & $\ast$ B4 & $2.983 \times 10^{31}$ (78) & $5.426 \times 10^{9}$ (78) & $15.12$ (74) & 27700 (78) & $310$ (74) & 5  \\
$\tau$ Sco & $\ast$ B0 V & $2.983 \times 10^{31}$ (78) & $4.174 \times 10^{9}$ (78) & $984.72$ (74) & 30000 (78) & $90$ (74) & 5\\
\hline
\end{tabular}
}
\caption{
Mass $M$, radius $R$, period $P$, effective temperature $T_{eff}$,
and surface magnetic field $B$ for
the objects in this study. The sources for these values are
indicated in parentheses with the associated references given in
the main text of the Appendix.
Footnotes:
a: Typical brown dwarf radius of $0.9 R_{\rm Jup}$ assumed
\citep{vrba04,kao16};
b: Mass derived from spectral type using the table in the indicated reference;
c: Maximum rotational period estimated in indicated reference;
f: Magnetic field values are based on different observational methods
listed in the last column with the individual technique given in the
main text of the Appendix.
}
\end{table*}

\section{Further details of observations}
\subsection{Data and error analysis}

\label{a:data_analysis}

\begin{table*}
      \caption[]{Details of the data analysis. Row information for
      calculation of fluxes and S/N.}
\label{table:data_analysis}      
\centering          
\begin{tabular}{l c c c c c c c}     
\hline\hline
 Type & $y_r$  & $n_1$ & $n_2$ & $a_1$ & $a_2$ & $b_1$ & $b_2$ \\ 
\hline
NUV         & 604 & -5 & 5 & 6 & 26 &-26 & -6\\
FUV         & 476 & -5 & 5 & 6 & 46 &-26 & -6\\
Ly-$\alpha$ & 486 & -5 & 5 & 6 & 36 &-36 & -6\\
\hline                  
\end{tabular}
\end{table*}

For the analysis of the observations we primarily worked with the highest
calibration level of STIS data provided within the {\it x2d}-files where spectral
energy fluxes in erg s$^{-1}$ cm$^{-2}$ $\AA^{-1}$ are given. Because
the target is extremely faint in the UV, the position of the target
within the slit could not be directly determined. We therefore used
the reference position provided in the headers of the data files
$y_{\mathrm{ref, nom}}$. If integrated fluxes along the trace (i.e., the direction of dispersion) in
individual or combined exposures show locally enhanced fluxes near
the expected location of the target, we worked with the improved
position. We neglected pixels on the detector where data quality flags
indicate quality issues for a given exposure and subsequently
renormalized integrated fluxes considering the missing pixels. 

Similar to the analysis in \citet{saur18}, the spectral flux
as a function of wavelength $\lambda$ represented by 
column $i_x$ is given by
\begin{eqnarray} 
f_{trace}(\lambda)  =\sum_{i_y=y_{ref}- n_1
}^{y_{ref}+ n_2 }  f(\lambda, i_y) \;. 
\end{eqnarray}
We calculate the average background flux per pixel $f^{px}_{bg}$ in
rows sufficiently above and below the rows where we expect
flux from the target
 with
\begin{eqnarray} 
f^{px}_{bg}(\lambda) =
\frac{
\sum_{i_y=y_{ref}+ a_1
}^{y_{ref}+ a_2 }  f(\lambda, i_y)  + \sum_{i_y=y_{ref}- b_1
}^{y_{ref}- b_2 }  f(\lambda, i_y)
}
{
a_2-a_1+1+b_2-b_1+1
}
,\end{eqnarray} 
with the positive integer numbers $n_1, n_2, a_1, a_2,
b_1, b_2$. The values of these integers depend on the 
type of aperture and grating for each exposure (see Table
\ref{table:data_analysis}).
In these calculations 
a separate background is calculated for
each column $i_x$ because the
background fluxes change along the dispersion axis.
The net flux from the target as a function of wavelength is thus given by
\begin{eqnarray}
f_{net}(\lambda) = f_{trace}(\lambda) - 
f^{px}_{bg}(\lambda)\left(
n_2-n_1+1
\right)
\;.
\label{e:net}
\end{eqnarray}

We calculate the variances and the  S/N
 in
two ways. The variances of the background for a certain wavelength
$\lambda$ represented by an individual column
$i_x$ 
is given by
\begin{eqnarray}
V^{px}(i_x) &= &
\frac{
\sum_{i_y=y_{ref}+ a_1
}^{y_{ref}+a_2 }  (f(i_x, i_y) -f^{px}_{bg}(i_x))^2+ \sum_{y_{ref}- b_1
}^{y_{ref}-b_2 }  (f(i_x, i_y) -f^{px}_{bg}(i_x))^2 
}
{
a_2-a_1+b_2-b_1+1
}
\;.
\label{e:V}
\end{eqnarray}
For comparison of net fluxes $f_{net}$ from the target with flux uncertainties,
the variance in each pixel $V^{px}$ needs to be normalized to the
number of rows used to calculate $f_{net}$.
For the calculation of  net  fluxes within 
certain wavelength ranges, the variances of the
individual columns within the selected wavelength
ranges  need to be summarized to the total variance $V$. 
The S/N values are finally calculated with the standard deviation
$\sigma=\sqrt{V}$.

Alternatively, we calculate the S/N based on the counts that
contribute to the calculation of certain fluxes (based on {\it
x2d}-files or  {\it flt}-files) using
\begin{eqnarray}
S/N = \frac{S-B}{\sqrt{S+B}} \;,
\label{e:SNR}
\end{eqnarray}
with the counts of the total signal $S$ and the background $B$. The
signal and the background are taken from the same pixels as described
in Sects. (\ref{e:net}) and (\ref{e:V}) and Table \ref{table:data_analysis}.

\subsection{Acquisition of target}
\label{a:acq}
   \begin{figure*}
   \centering
   \includegraphics[width=14.cm]{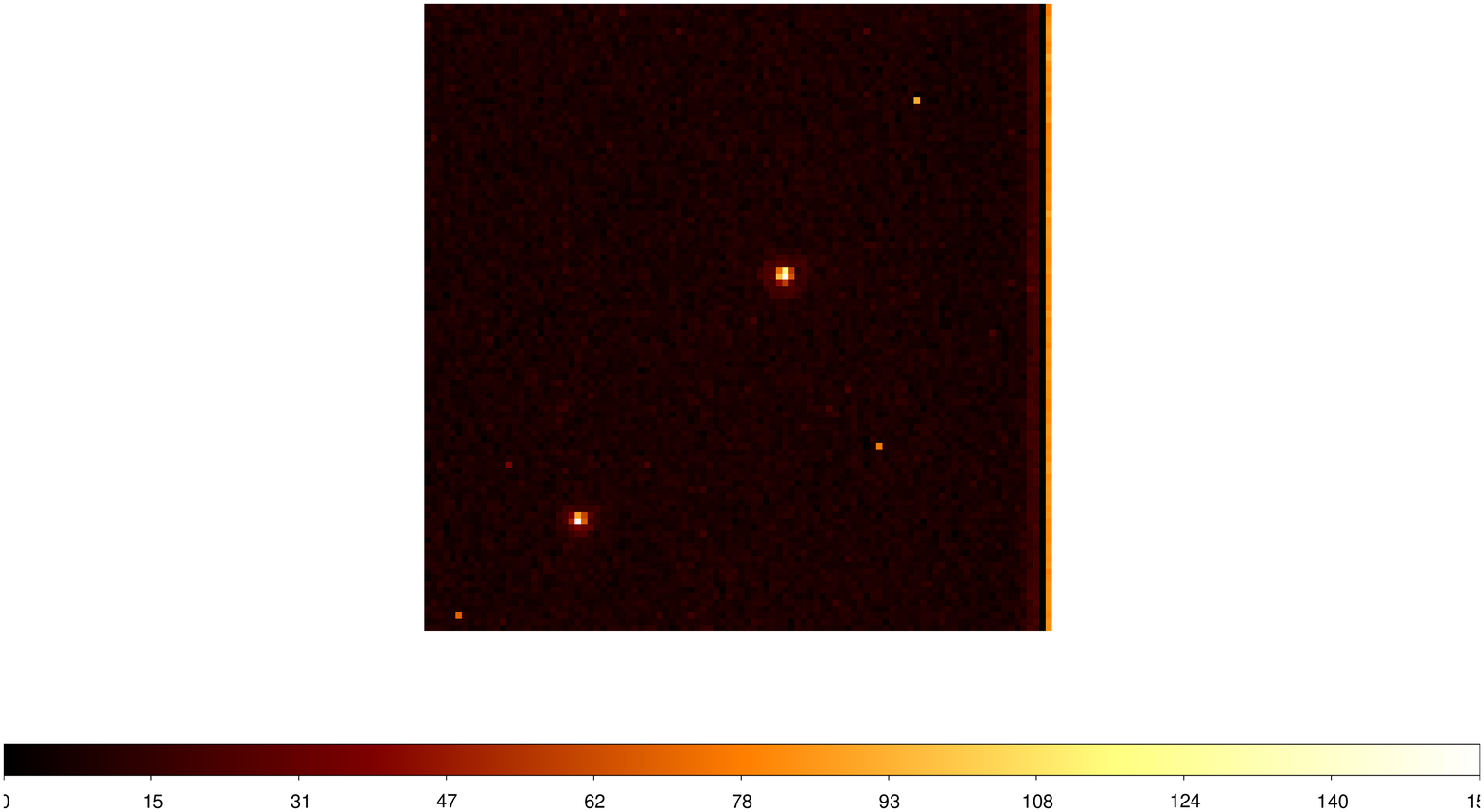}
   \caption{Acquisition image}
   \label{f:acq1}%
    \end{figure*}
Figure \ref{f:acq1} shows an image of the first acquisition exposure
of visit 1 blind-pointing into an area of 5 by 5 arcsec.
The brightest object in Figure
\ref{f:acq1} is located at RA = 12:37:35.7807 and DEC =
65:26:03.250.
The standard deviations for the expected positional accuracy in right
ascension and declination of the target in the acquisition image has been
determined as the root sum squared of the standard deviations of the initial
position of 2MASS J1237+6526 in the 2MASS catalogue (0.19", 0.18"),
of the proper motion correction (0.17", 0.12") \citep{vrba04},
and of the pointing accuracy of HST using stars in the guide star catalogue
(0.2", 0.2") as (0.32", 0.30").
The offset of the brightest object from the expected position is
(-0.07", 0.47"), hence we derive a total positional offset of 1.6 $\sigma$.
The brightness of the
object in the acquisition image corresponds to 43\% of the expected
brightness of the target. This estimate stems from runs with the STIS
exposure time calculator using the spectrum of 2MASS J1237+6526 from
\citet{burg02} and \citet{lieb07}.

Separated by 2.6'', there is a second slightly dimmer object in the acquisition
image. Its location is 7.0 $\sigma$ away from the expected location of
2MASS J1237+6526 and its total counts in the acquisition exposure amount
to 34\% of the counts expected from 2MASS J1237+6526. In particular
due to the significant deviation from the expected position, we do
not expect the second object to be our desired target. Unfortunately,
we could not identify the second object within standard data catalogues. 

The subsequent acquisition procedure by HST positioned the brightest
object of the initial acquisition exposure into the desired slits of
the science exposures. The acquisition image of visit 2 is
basically identical.

\end{appendix}



%
%

\end{document}